\documentclass[prb,twocolumn,superscriptaddress,citeautoscript,showpacs,amsart]{revtex4}

\usepackage{graphicx}
\usepackage{multirow}
\usepackage{color}
\usepackage{bm}
\usepackage{times}
\usepackage{amsmath,bm,amsfonts}
\usepackage{dcolumn}
\usepackage{graphicx}
\usepackage{latexsym}
\usepackage{physics}
\usepackage{ulem} % to strike things out \normalem % usual emph
\usepackage{braket}

\def\ket#1{|#1\rangle }
\def\bra#1{\langle #1 |}
\def\n{\nonumber \\ }

\newcommand{\f}{\frac}

\newcommand{\V}{\vec}
\newcommand{\B}{\beta}

\begin{document}

\title{Magnetic field induced topological semimetals near a quantum critical point of pyrochlore iridates}

\author{Taekoo \surname{Oh}}
\affiliation{Department of Physics and Astronomy, Seoul National University, Seoul 08826, Korea}

\affiliation{Center for Correlated Electron Systems, Institute for Basic Science (IBS), Seoul 08826, Korea}

\affiliation{Center for Theoretical Physics (CTP), Seoul National University, Seoul 08826, Korea}

\author{Hiroaki \surname{Ishizuka}}
\affiliation{Department of Applied Physics, The University of Tokyo, Hongo, Bunkyo-ku, Tokyo 113-8656, Japan}

\author{Bohm-Jung \surname{Yang}}
\email{bjyang@snu.ac.kr}
\affiliation{Department of Physics and Astronomy, Seoul National University, Seoul 08826, Korea}

\affiliation{Center for Correlated Electron Systems, Institute for Basic Science (IBS), Seoul 08826, Korea}

\affiliation{Center for Theoretical Physics (CTP), Seoul National University, Seoul 08826, Korea}

\date{\today}

\begin{abstract}
Motivated by the recent experimental observation of anomalous magneto-transport properties near the Mott quantum critical point (QCP) of pyrochlore iridates, we study the generic topological band structure near QCP in the presence of magnetic field.
We have found that the competition between different energy scales can generate various topological semi-metal phases near QCP. Here the central role is played by the presence of a quadratic band crossing (QBC) with four-fold degeneracy in the paramagnetic band structure. Due to the large band degeneracy and strong spin-orbit coupling, the degenerate states at QBC can show an anisotropic Zeeman effect as well as the conventional isotropic Zeeman effect. Through the competition between three different magnetic energy scales including the exchange energy between Ir electrons and two Zeeman energies, various topological semimetals can be generated near QCP. Moreover, we have shown that these three magnetic energy scales can be controlled by modulating the magnetic multipole moment (MMM) of the cluster of spins in a unit cell, which can couple to the intrinsic MMM of the degenerate states at QBC.
We propose the general topological band structure under magnetic field achievable near QCP, which would facilitate the experimental discovery of novel topological semimetal states in pyrochlore iridates. 
\end{abstract}

\pacs{}

\maketitle
 
% **************************************** Introduction ****************************************** %
\section{Introduction}

\begin{figure}[t]
\centering
\includegraphics[width=\columnwidth]{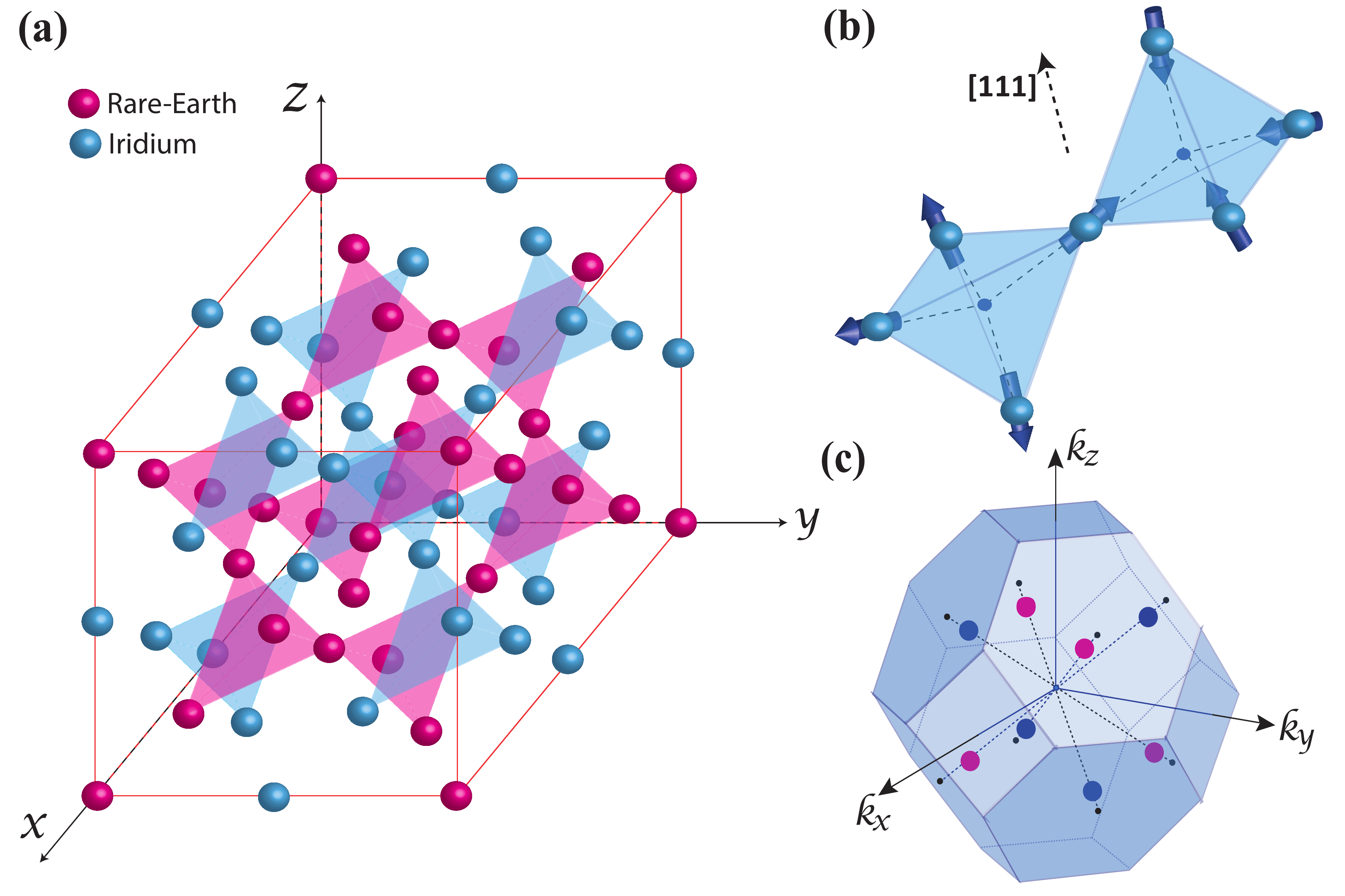}
\caption{ 
(a) The lattice structure of pyrochlore iridates.
(b) All-in all-out (AIAO) magnetic ordering.
(c) Distribution of Weyl points in the Weyl semimetal related with AIAO ordering.
}
\label{fig:1}
\end{figure}

Electron correlation and spin-orbit coupling are two quintessential ingredients underlying vast emergent physical phenomena in condensed matters\cite{witczak2014correlated,schaffer2016recent}.
In particular, when these two energy scales are comparable to the electron bandwidth, various correlated phases with novel topological properties are expected to appear in general\cite{wan2011topological,lv2015experimental,weng2015weyl,huang2015weyl,xu2015discovery}. Pyrochlore iridates with the chemical formula R$_2$Ir$_2$O$_7$ (R: a rare earth ion, see Fig.~\ref{fig:1}(a)) are a representative example of such correlated topological systems that can potentially host various intriguing electronic states\cite{witczak2014correlated,schaffer2016recent}. 
In the paramagnetic metal (PM) phase, it was theoretically predicted that these materials have a quadratic band crossing (QBC) with doubly-degenerate hole-like and electron-like bands touching at the $\Gamma$ point\cite{witczak2013pyrochlore}.
Recent ARPES study on Pr$_2$Ir$_2$O$_7$\cite{EGM2015NaCom} finds electron dispersion which conforms closely to this prediction.
When a magnetic transition occurs below the temperature $T_{N}$, a variety of interesting electronic states possibly show up from the QBC. For instance, an antiferromagnetic (AF) Weyl semimetal (WSM) phase is theoretically predicted to exist between a PM and an AF insulator (AFI) with all-in all-out (AIAO) type magnetic ordering shown in Fig.~\ref{fig:1}(b,c)\cite{wan2011topological,kurita2011topological,tomiyasu2012emergence,witczak2012topological,witczak2013pyrochlore}.

%%%%%%%%%%%%%%%%%%%%%%%%%%%%%%%%%%%%%%%%%%%%%%%%%%%%
%%%%%%%%%%%%%%%%%%%%%%%%%%%%%%%%%%%%%%%%%%%%%%%%%%%%%

%%%%%%%%%%%%%%%%%%%%%%%%%%%%%%%%%%%%%%%%%%%%%%%%%%%%
\begin{figure}[t]
\centering
\includegraphics[width=0.75\columnwidth]{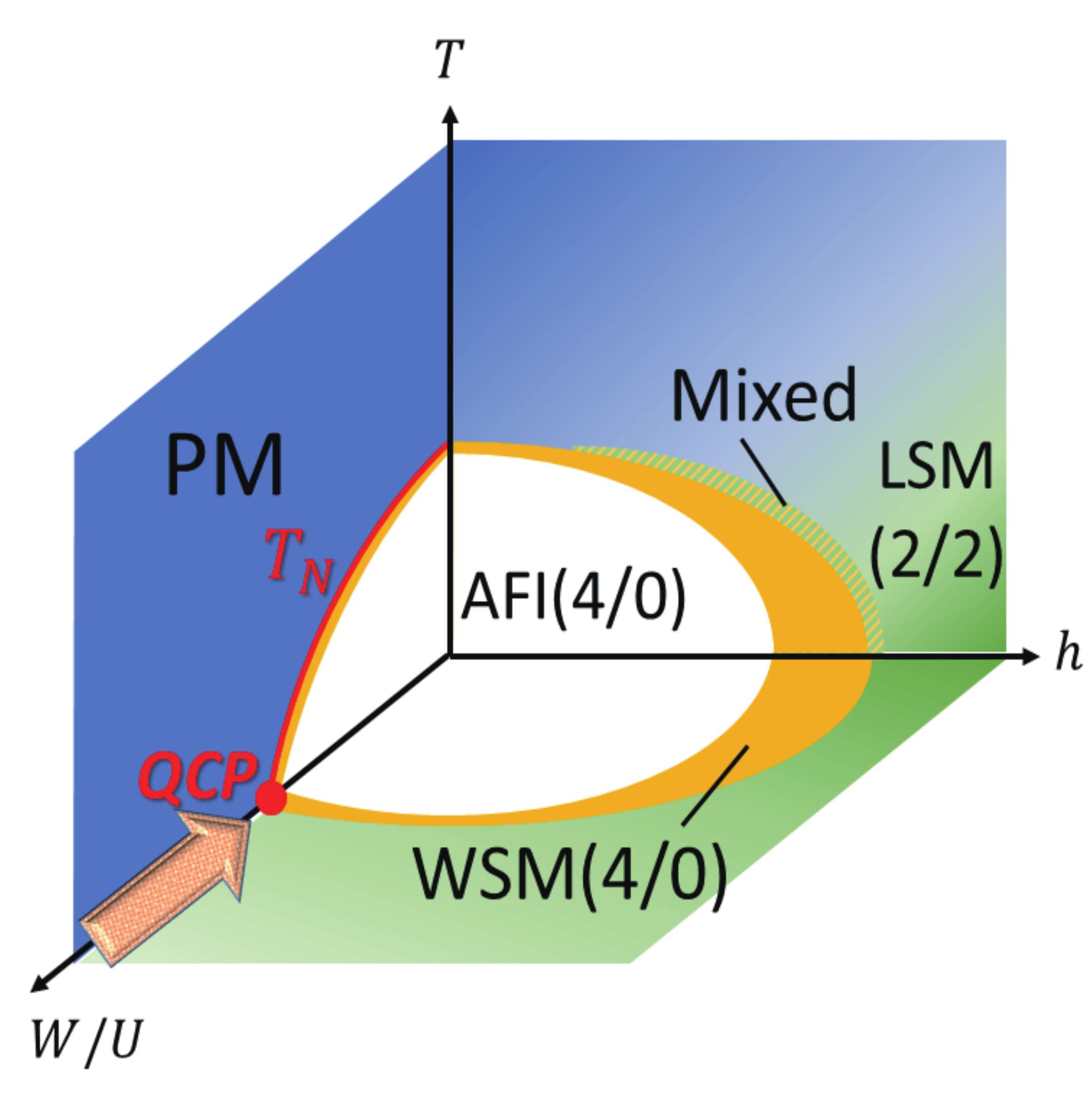}
\caption{ 
Schematic phase diagram near the quantum critical point (QCP) in the space of Coulomb interaction $U$, temperature $T$, and magnetic field $h$ for given electron bandwidth.
PM is a paramagnetic metal, AFI(4/0) is an antiferromagnetic insulator with AIAO, WSM(4/0) is Weyl semimetal with AIAO, and LSM(2/2) is a line-node semimetal with 2-in-2-out ordering.
}
\label{fig:2}
\end{figure}
%%%%%%%%%%%%%%%%%%%%%%%%%%%%%%%%%%%%%%%%%%%%%%%%%%%%%

On the other hand, in reality, except the case of R=Pr where PM phase persists down to the lowest temperature accessible, the WSM state only appears in a small window at the boundary between PM and AFI phases [Fig.~\ref{fig:2}].
However, by substituting R sites by the ions with larger radius or applying hydrostatic pressure, one can reduce $T_{N}$ systematically and approach the quantum critical point (QCP), around which a semimetallic ground state with AF ordering may be achievable\cite{ueda2017magnetic}.
Interestingly, in systems close to the QCP such as those with R=Nd or Pr, anomalous transport properties are observed such as anomalous Hall effects, metallic states at AIAO domain walls, magnetic field induced metal-insulator transitions, etc\cite{machida2007unconventional,balicas2011anisotropic,disseler2013magnetization,ueda2014anomalous,ueda2015magnetic,ma2015mobile,tian2016field, goswami2017competing,chen2012magnetic}. 
In particular, a recent study of (Nd$_{1-x}$Pr$_x$)$_{2}$Ir$_{2}$O$_{7}$ under pressure in which $T_{N}$ has been systematically tuned to reach the QCP, has demonstrated unusual magnetotransport properties near the QCP, which might be associated with topological semimetal phases emerging near the QCP under magnetic field\cite{ueda2017magnetic}.
The accumulated experimental and theoretical results from preceding studies are summarized in the schematic phase diagram shown in Fig.~\ref{fig:2}, implying that applying magnetic field to the system located near the QCP is a promising way to achieve various topological semimetals with point or line nodes.

The main purpose of the present theoretical study is to provide a general theoretical framework to understand the magnetic field induced topological semimetals emerging near the QCP of pyrochlore iridates. 
To address this issue, we start from the PM phase with QBC and approach the QCP by introducing AIAO ordering together with magnetic field.
The QBC at the $\Gamma$ point can be described by the states carrying the total angular momentum $J=3/2$. 
Due to the large total angular momentum $J$ and strong spin-orbit coupling, the Zeeman coupling shows a non-trivial feature; the Zeeman field $\vec{H}$ can give rise to an unconventional anisotropic Zeeman effect ($\propto \vec{H}\cdot\vec{J}^{3}$) as well as the usual isotropic Zeeman coupling ($\propto \vec{H}\cdot\vec{J}$).
Moreover, an additional magnetic energy scale associated with the AIAO ordering exists.
Since the exchange energy associated with AIAO ordering and the two different Zeeman energies are comparable near the QCP, the competition between them can bring about various novel topological semimetal phases according to the low energy theory.
In terms of microscopic lattice degrees of freedom, we show that the interplay between three different magnetic energy scales can be compactly described in terms of magnetic multipole moments (MMM) of the cluster of four spins in a tetrahedron.
Magnetic field induced modulation of MMM of the unit cell and its coupling to the intrinsic MMM of the degenerate states at QBC, lie at the heart of emergent topological semimetals near the QCP of pyrochlore iridates under magnetic field.

The paper is organized as follows. In Sec. II, we first introduce the effective theory at $\Gamma$ point, and describe topological semimetals induced by AIAO ordering. Magnetic-field induced topological semimetals are described by considering Zeeman field as well as AIAO ordering in Sec. III. In Sec. IV, we study the lattice model, and explain its relation with effective Hamiltonian analysis in terms of cluster magnetic multipole moments (CMMM). At last, in Sec. V, we conclude.

%%%%%%%%%%%%%%%%%%%%%%%%%%%%%%%%%%%%%%%%%
\begin{figure}[b]
\centering
\includegraphics[width=\columnwidth]{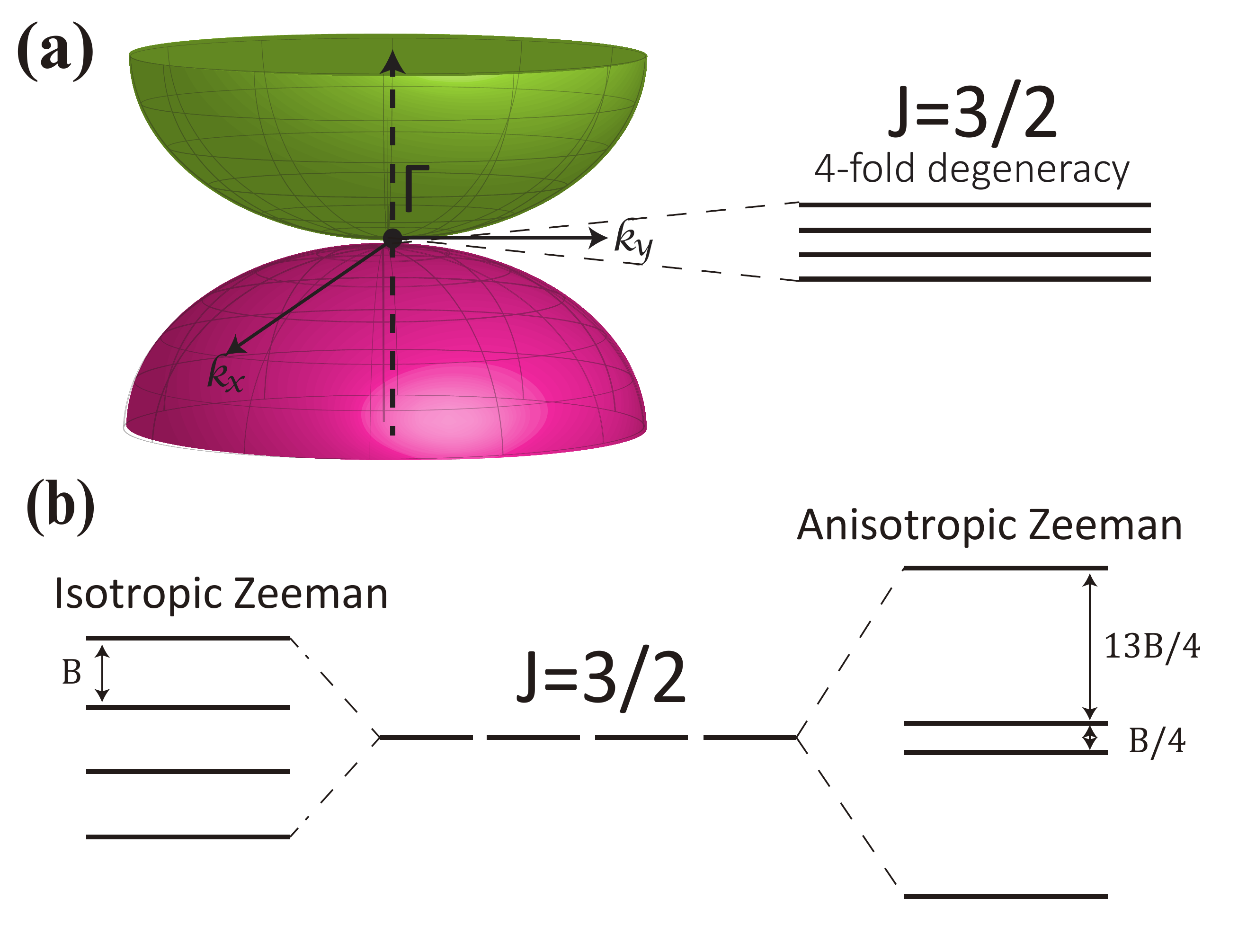}
\caption{(a) The band structure near the quadratic band crossing (QBC).
(b) The energy level splitting of $J=3/2$ states at $\Gamma$ point due to isotropic and anisotropic Zeeman terms, respectively.
}
\label{fig:3}
\end{figure}
%%%%%%%%%%%%%%%%%%%%%%%%%%%%%%%%%%%%%%%%

\section{Quadratic Band Crossing and AIAO ordering}

The QBC of the PM state\cite{witczak2012topological,witczak2013pyrochlore,EGM2015NaCom,cano2017chiral} 
is shown in Fig.~\ref{fig:3}(a).
Since each eigenstate is doubly degenerate due to the time-reversal and inversion symmetries, the QBC at $\Gamma$ has four-fold degeneracy with the total angular momentum $J=3/2$.
The low energy physics near the QBC can be described by the so-called Luttinger Hamiltonian\cite{luttinger1956quantum} given by
\begin{align}
\mathcal{H}_0(\vec k) = \epsilon_0(\vec k) + \sum_{i=1}^5 d_i(\vec k) \Gamma_i, 
\label{eq:Luttinger}
\end{align}
where $\epsilon_0(\vec k) = k^2/2m$ and $\Gamma_{i}$ is a 4$\times$4 gamma matrix satisfying the Clifford algebra $\{\Gamma_{i},\Gamma_{j}\}=2\delta_{ij}$ ($i,j=1\sim 5$.). 
By defining ten additional Hermitian matrices as $\Gamma_{ij}=[\Gamma_i,\Gamma_j]/2i$ and the identity matrix, one can find a complete set of sixteen Hermitian $4\times 4$ matrices.
The detailed form of the function $d_{1\sim 5}(\vec{k})$ constrained by the cubic symmetry at $\Gamma$, is shown in \ref{sec:App1}. 

When Ir AIAO ordering is developed below $T_{N}$, the QBC at $\Gamma$ splits into four pairs of Weyl points (WPs) in which each pair is aligned along either [111] or its three other symmetry-related directions\cite{witczak2012topological}.
Such an emerging WSM with eight WPs can be described by adding $\mathcal{H}_{\text{AIAO}}=-\alpha\Gamma_{45}$ with $\alpha\propto Um_{\text{AIAO}}$ to Eq.~(\ref{eq:Luttinger}) where $U$ is the local Coulomb repulsion and $m_{\text{AIAO}}$ represents the local magnetic moment of the AIAO state.
Since the separation between the WP pair on the [111] axis is proportional to $\sqrt{|\alpha|}$, when the $\alpha$ becomes bigger than the critical value $\alpha_{c}$ at which WP pairs hit the Brillouin zone boundary and pair-annihilate, the system becomes a gapped insulator.
According to the previous theoretical study\cite{witczak2013pyrochlore}, such a pair-creation and pair-annihilation processes can be completed only within one-percent variation of $U/t$ ratio, where $t$ is the nearest neighbor hopping amplitude.
Thus the WSM phase can occupy a very narrow region of the phase diagram, which reflects the difficulty in approaching it in experiment.

%%%%%%%%%%%%%%%%%%%%%%%%%%%%%%%%%%%%%%%%%
\begin{figure}[t]
\centering
\includegraphics[width=\columnwidth]{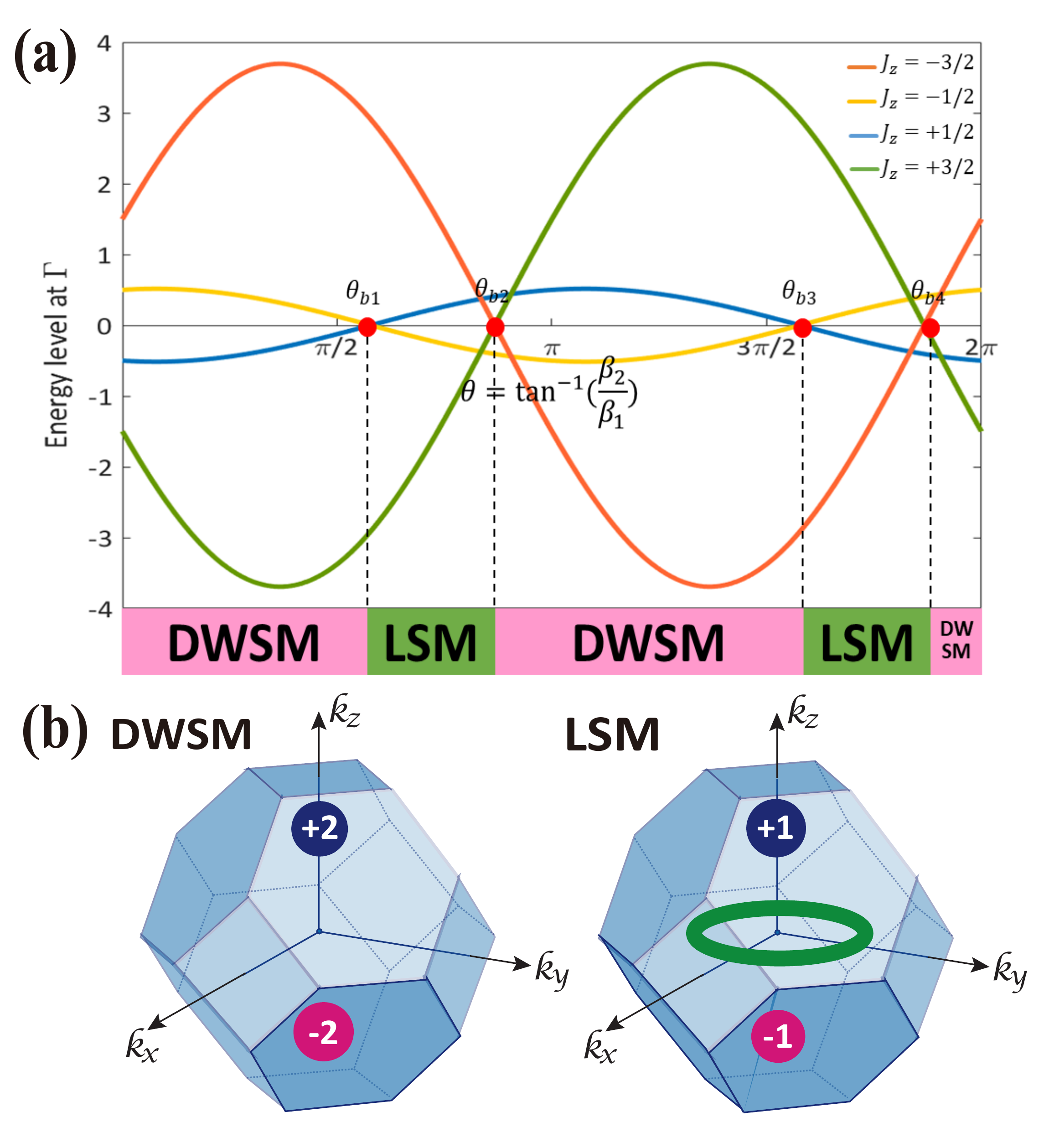}
\caption{
(a) Change of energy levels at $\Gamma$ point when $\theta=\tan^{-1}[\beta_{2}/\beta_{1}]$ varies, and the associated phase diagram. Here we assume $\hbar=1$ and plot the energy per unit magnetic field $E/B$. DWSM (LSM) denotes a double Weyl semimetal (a line-node semimetal). 
(b) Distribution of point/line nodes in DWSM and LSM.
}
\label{fig:4}
\end{figure}
%%%%%%%%%%%%%%%%%%%%%%%%%%%%%%%%%%%%%%%%

\section{Topological semimetals induced by Zeeman field}

On the other hand, when magnetic field is applied to the semimetal with QBC, various topological semimetals can emerge.
The influence of the external Zeeman field $\vec{H}$ on QBC can be described by 
\begin{align}
\mathcal{H}_{B}=-\beta_1 \vec B \cdot \vec J - \beta_2 \vec B \cdot \vec J^3, \label{eq:Zeemanterms}
\end{align} 
where $\vec J = (J_x, J_y, J_z)$, $\vec J^3 = (J_x^3, J_y^3,J_z^3)$, and $\vec B=\vec{B}(\vec{H},\vec{M},...)$ indicates the effective Zeeman field including $\vec{H}$ and the average magnetization $\vec{M}$.
Two constants $\beta_{1}$ and $\beta_{2}$ measure the magnitude of the isotropic and anisotropic Zeeman terms, respectively.
The anisotropic Zeeman term coupled with the cubic invariant $\vec{J^3}$ arises due to spin-orbit coupling and the large total angular momentum $J=3/2$.
Normally, the anisotropic Zeeman term, that has been known as the q-term in the Luttinger Hamiltonian, is proportional to spin-orbit coupling and makes a tiny contribution to Zeeman splitting\cite{hensel1969anisotropy, nenashev2003wave}.
However, in pyrochlore iridates, it can make a significant contribution to the energy splitting at the $\Gamma$ point whose magnitude can even be controlled by modulating the orientation of spins within a unit cell as explained below.

%%%%%%%%%%%%%%%%%%%%%%%%%%%%%%%%%%%%%%%%%%%%%%%%%%%%
\begin{figure}[b]
\centering
\includegraphics[width=\columnwidth]{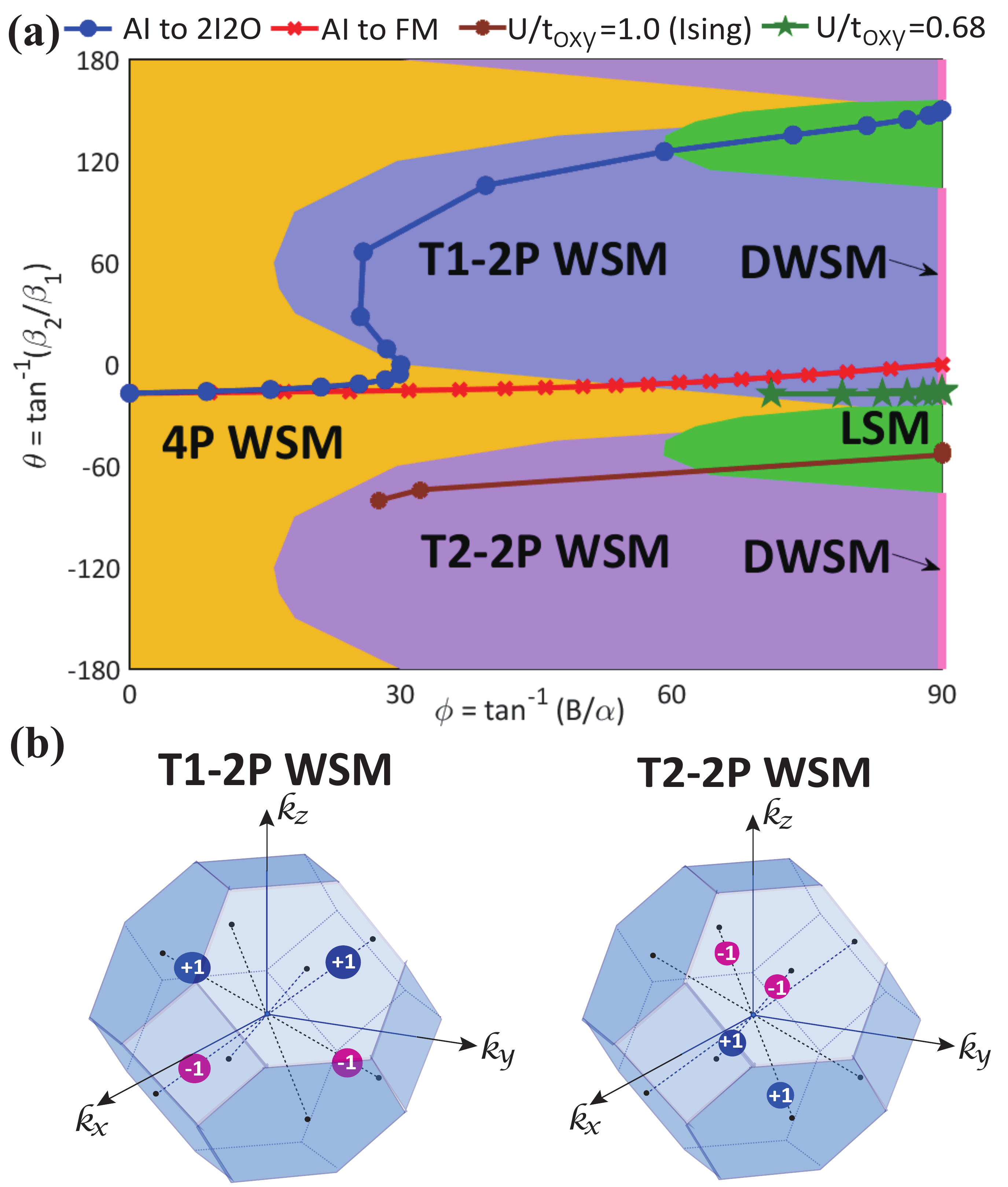}
\caption{
(a) General phase diagram under magnetic field along $[001]$ in the $(\theta, \phi)$ plane obtained from the extended Luttinger model. Here 4P WSM indicates a WSM
with four pairs of WPs whereas T1-2P WSM (T2-2P WSM) denotes type-1 (type-2) WSM with two pairs of WPs. 
The solid lines indicate the trajectory followed by the mean-field lattice model when the spin orientation in a unit cell changes.
(b) Distribution of WPs in T1/T2-2P WSM.
}
\label{fig:5}
\end{figure}
%%%%%%%%%%%%%%%%%%%%%%%%%%%%%%%%%%%%%%%%%%%%%%%%%%%

In general, the isotropic Zeeman term splits the degenerate eigenstates carrying different $J_{z}$, leading to equally spaced energy levels at the $\Gamma$ point as shown in Fig.~\ref{fig:3}(b).
Thus in systems with $\beta_{1}\neq 0$ and $\beta_{2}=0$, Zeeman field $\vec{H}$ cannot make a level crossing between the states with different $J_{z}$ at the $\Gamma$ point.
On the other hand, when the isotropic and anisotropic Zeeman terms exist simultaneously, the energy ordering between states with different $J_{z}$ can be rearranged depending on the ratio $\beta_{2}/\beta_{1}=\tan\theta$.
Fig.~\ref{fig:4}(a) shows the evolution of energy levels at the $\Gamma$ point as $\theta$ varies when $\vec{H}\parallel[001]$. 
One can clearly see the level crossing at several critical angles $\theta_{c}$ which indicates topological phase transitions between different topological semimetals.
As shown in Fig.~\ref{fig:4}(a), when $\vec{H}\parallel[001]$, one can obtain either a double Weyl semimetal (DWSM) having two WP with the monopole charge $\pm2$ on the $k_{z}$ axis or a line-node semimetal (LSM) having a circular nodal line on the $k_{z}=0$ plane with two additional WP on the $k_{z}$ axis.
On the other hand, when $\vec{H}\parallel[111]$, since the residual symmetry of the system is lower than the case with $\vec{H}\parallel[001]$, band crossing at $\Gamma$ can occur in a more limited situation, thus the resulting topological phase diagram is simpler as detailed in \ref{sec:App1}.

When magnetic field is applied to real materials, both AIAO ordering and two Zeeman terms exist simultaneously.
Then the most general low energy band structure can be captured by the extended Luttinger model $\mathcal{H}_{\text{extended}}\equiv\mathcal{H}_{0}+\mathcal{H}_{\text{AIAO}}+\mathcal{H}_{\text{B}}$.
Since there are three competing energy scales $\alpha,~\beta_{1},~\beta_{2}$, one can obtain the general phase diagram in the two-dimensional $(\theta,\phi)$ plane where the angular variable $\phi\equiv\tan^{-1}(B/\alpha)$ is introduced to measure the importance of the Zeeman term relative to the energy scale for the AIAO ordering. 
As shown in Fig.~\ref{fig:5}(a), various novel topological semimetal phases can arise by tuning $\theta$ and $\phi$.

\section{Lattice model and cluster magnetic multipole moments (CMMM)}

To provide a microscopic picture for magnetic-field induced topological semimetals in lattice systems, we study a tight-binding Hamiltonian $H=H_{0}+H_{U}+H_{Z}+H_{fd}$, where $H_U =U \sum_{i} n_{i\uparrow} n_{i\downarrow}$ is the on-site Hubbard interaction, $H_{Z} = \sum_{i,s} c_{i,s}^\dagger\frac{ (\vec H \cdot \vec\sigma_{ss'})}{2}c_{i,s'}$ indicates the Zeeman coupling, and $H_{fd}$ denotes the exchange coupling between Ir and Nd moments.
$c_{i,s}$ ($c_{i,s}^\dagger$) is the annihilation (creation) operator for electrons carrying spin $s=\uparrow,\downarrow$ on $i$th site, $n_{is}=c_{i,s}^\dagger c_{i,s}$ is the electron number opertor.
Here it is assumed that each Ir ion carries an effective spin 1/2 moment represented by the Pauli matrix $\vec{\sigma}$. The hopping process between Ir sites is described by
\begin{align}
H_0 =& \sum_{s,s'}[\sum_{\langle ij \rangle} c_{i,s}^\dagger (t_1 + it_2 \vec d_{ij} \cdot \vec \sigma_{ss'}) c_{j,s'} \n
&+ \sum_{\langle \langle ij \rangle \rangle} c_{i,s}^\dagger(t_1' + i\{t_2'\vec R_{ij} + t_3' \vec D_{ij}\}\cdot \vec \sigma_{ss'}) c_{j,s'}], \label{eq:TB}
\end{align}
where $t_{1}$ ($t_{1}'$) denotes the spin-independent hopping amplitude between nearest-neighbor (next-nearest-neighbor) sites, and $t_{2}$, $t_{2,3}'$ indicate spin-dependent hopping amplitudes including the oxygen mediated hopping amplitude $t_{oxy}$ as well as the direct hopping amplitudes between Ir ions~\cite{witczak2012topological,witczak2013pyrochlore}.
The Hubbard interaction term is treated by a mean field theory ($H_{U}\approx H_{U}^{\text{MF}}$) by introducing local order parameters $\vec{m}_{\alpha}\equiv\frac{1}{2N}\sum_{\bm{k}}\langle c^{\dag}_{\alpha,s}(\bm{k})\vec{\sigma}_{s,s'}c_{\alpha,s'}(\bm{k})\rangle$ where $\alpha=1,2,3,4$ indicates the four spins within a unit cell.
For $H_{fd}$, Nd moments are treated classically. (See \ref{sec:App4}.)

\begin{figure}[t]
\centering
\includegraphics[width=\columnwidth]{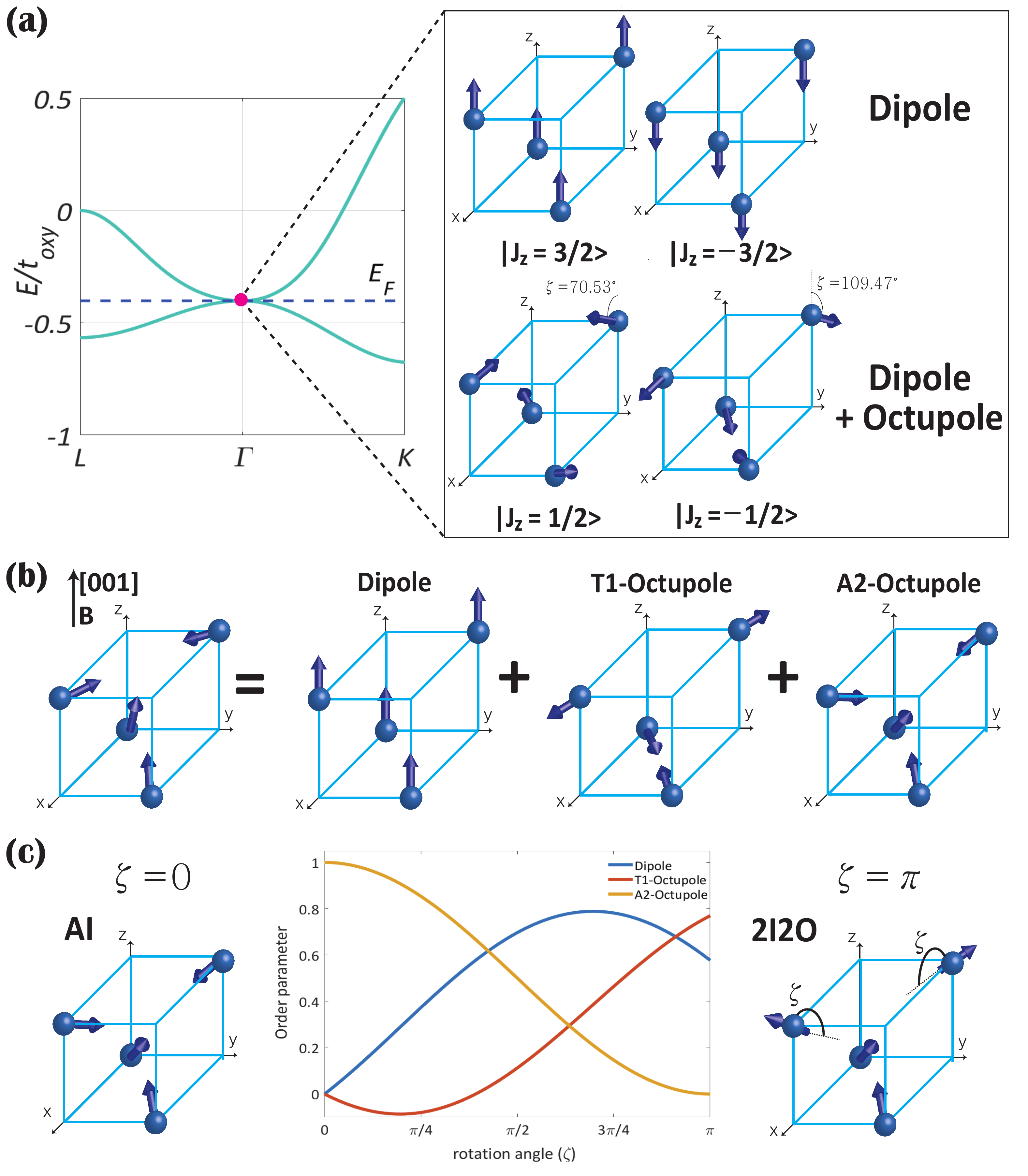}
\caption{
(a) Description of the wave functions for the four degenerate states at QBC.
(b) Decomposition of a generic spin configuration of a unit cell under $H \parallel [001]$ into magnetic multipole components. 
(c) Variation in the amplitudes of each multipolar components as the spin structure in a unit cell
is modulated from AI to 2I2O form under $H \parallel [001]$. 
}
\label{fig:6}
\end{figure}

Fig.~\ref{fig:6}(a) shows the band structure of PM obtained by solving $H_{0}$.
One can clearly see the presence of a QBC at the $\Gamma$ point that can be effectively described by the Luttinger Hamiltonian discussed before.
To understand the nature of the four degenerate states at the QBC carrying $J=3/2$, we have depicted the relevant wave functions in Fig.~\ref{fig:6}(a). 
One intriguing property of these degenerate eigenstates is that they intrinsically carry cluster magnetic multipole moments (CMMM) defined below. Namely, the states with the angular momentum $J_{z}=\pm 3/2$ carry cluster magnetic dipole moments whereas the other two states with $J_{z}=\pm 1/2$ carry cluster magnetic dipole and octupole moments.
Due to this intrinsic CMMM, those four states can selectively couple to specific magnetic ordering patterns of a magnetically ordered phase.

The MMM for a cluster of atoms are recently introduced by Suzuki et al. in Ref.~\onlinecite{suzuki2017cluster}.
Analogous to the local multiple moment of an atom\cite{kusunose2008description}, the rank-$p$ MMM of a given cluster $\mu$ is defined as $M_{pq}^{\mu} = \sqrt{\frac{4\pi}{2p+1}} \sum_{i=1}^{N} \vec{m}_i \cdot \nabla_i (|R_i|^p Y_{pq}^{*}(\theta_i,\phi_i))$ where $q$ is the magnetic quantum number ranging from $-p$ to $p$, $N$ is the number of atoms in a cluster, $\vec{m}_i$ is the magnetic moment vector at the $i$-th atom of the cluster, $(R_i, \theta_i, \phi_i)$ is the spherical coordinate of $i$-th atom, and $Y_{pq}$ is the spherical harmonics.
By taking summation over all clusters in the magnetic unit cell, the $p$-th order of CMMM can be obtained. 

%%%%%%%%%%%%%%%%%%%%%%%%%%%%%%%%%%%%%%%%%%%%
%%%%%%%%%%%%%%%%%%%%%%%%%%%%%%%%%%%%%%%%%%%%%%%

The CMMM of a tetrahedral unit cell can be analyzed further as follows.
Counting the three components of a spin separately, the twelve independent spin degrees of freedom in a unit cell can be classified by using group theory.
The resulting symmetrized spin configuration with a fixed CMMM can be taken as a basis to represent the general spin configuration in a unit cell.
For instance, when $\vec{H}\parallel[001]$, the most general configuration of the four spins in a unit cell satisfying the lattice symmetry $C_{2z}$ and $\sigma_{d}T$ can be written as 
\begin{align}
|\psi\rangle_{[001]}=a_{D}|D\rangle+a_{T_{1}}|T_{1}\rangle+a_{A_{2}}|A_{2}\rangle,
\end{align} 
where $|D\rangle$, $|T_{1}\rangle$, $|A_{2}\rangle$ represent the basis states carrying cluster magnetic dipole, $T_{1}$-octupole, $A_{2}$-octupole moments, respectively, and $a_{D}$, $a_{T_{1}}$, $a_{A_{2}}$ represent the relevant amplitudes. (See Fig.~\ref{fig:6}(b).)
Changing the spin orientations,  $a_{D}$, $a_{T_{1}}$, $a_{A_{2}}$ can be tuned continuously as shown in Fig.~\ref{fig:6}(c).

Now let us describe how the intrinsic CMMMs of the four degenerate states at the QBC couple to the CMMM of a magnetically ordered phase.
To understand the relation between the CMMM of a lattice system and the three magnetic terms $\alpha$, $\beta_{1}$, $\beta_{2}$ of the extended Luttinger Hamiltonian, one can project the effective Zeeman term $H_{B}= \frac{1}{2}\sum_{i,s} \vec{B}_{\text{eff},i}\cdot \left[c_{i,s}^\dagger\vec\sigma_{ss'}c_{i,s'}\right]$ to the subspace spanned by the four degenerate states at QBC.
Here the local effective magnetic field $\vec{B}_{\text{eff},i}$ includes the influence of all interaction terms within the mean field theory, and should be determined self-consistently for given $\vec{H},~U,~J_{fd}$, and hopping parameters.
By using the projection operator $\hat{P}_{J}=\sum_{J_{z}}|J_{z}\rangle\langle J_{z}|$ where $|J_{z}\rangle$ indicates of the four degenerate states at QBC with the angular momentum $J_{z}$,
\begin{align}
\hat{P}_{J}H_{B}\hat{P}_{J} 
=M_{A_{2}}\Gamma_{45}+\left[\frac{2}{3}M_{D}-\frac{9}{4}M_{T_{1}}\right]J_{z}+M_{T_{1}}J_{z}^{3}, \label{eq:5}
\end{align}
for [001] field, where $M_{A_{2}}$, $M_{D}$, $M_{T_{1}}$ indicate the $A_{2}$ octupole moment (or AIAO order parameter), the magnetic dipole moment (or magnetization), the $T_{1}$ octupole moments, respectively.
It is worth to note that $M_{D}$ and $M_{T_{1}}$ determine the relative importance between the isotropic and anisotropic Zeeman terms. 
Since the CMMMs determine the three magnetic terms $\alpha$, $\beta_{1}$, $\beta_{2}$, one can expect that various topological semimetals predicted by the extended Luttinger model can be realized simply by changing the spin directions that controls the CMMMs.  

To demonstrate this idea, we have determined $\alpha$, $\beta_{1}$, $\beta_{2}$ by projecting the lattice model for various processes of changing spin orientations, and plotted the relevant trajectories in Fig.~\ref{fig:5}(a).
For instance, the red (blue) line in Fig.~\ref{fig:5}(a) describes the trajectory when the effective Zeeman field $\vec{B}_{\text{eff,i}}$ rotates the spins in a unit cell continuously from the AIAO configuration to the collinear ferromagnetic (2-in 2-out) state.
Depending on how the spin orientation changes, the CMMM of the unit cell and $\alpha$, $\beta_{1}$, $\beta_{2}$ change differently, which results in distinct trajectories and associated topological semimetals.

%%%%%%%%%%%%%%%%%%%%%%%%%%%%%%%%%%%%%%%%%%%%
\begin{figure}[t]
\centering
\includegraphics[width=0.8\columnwidth]{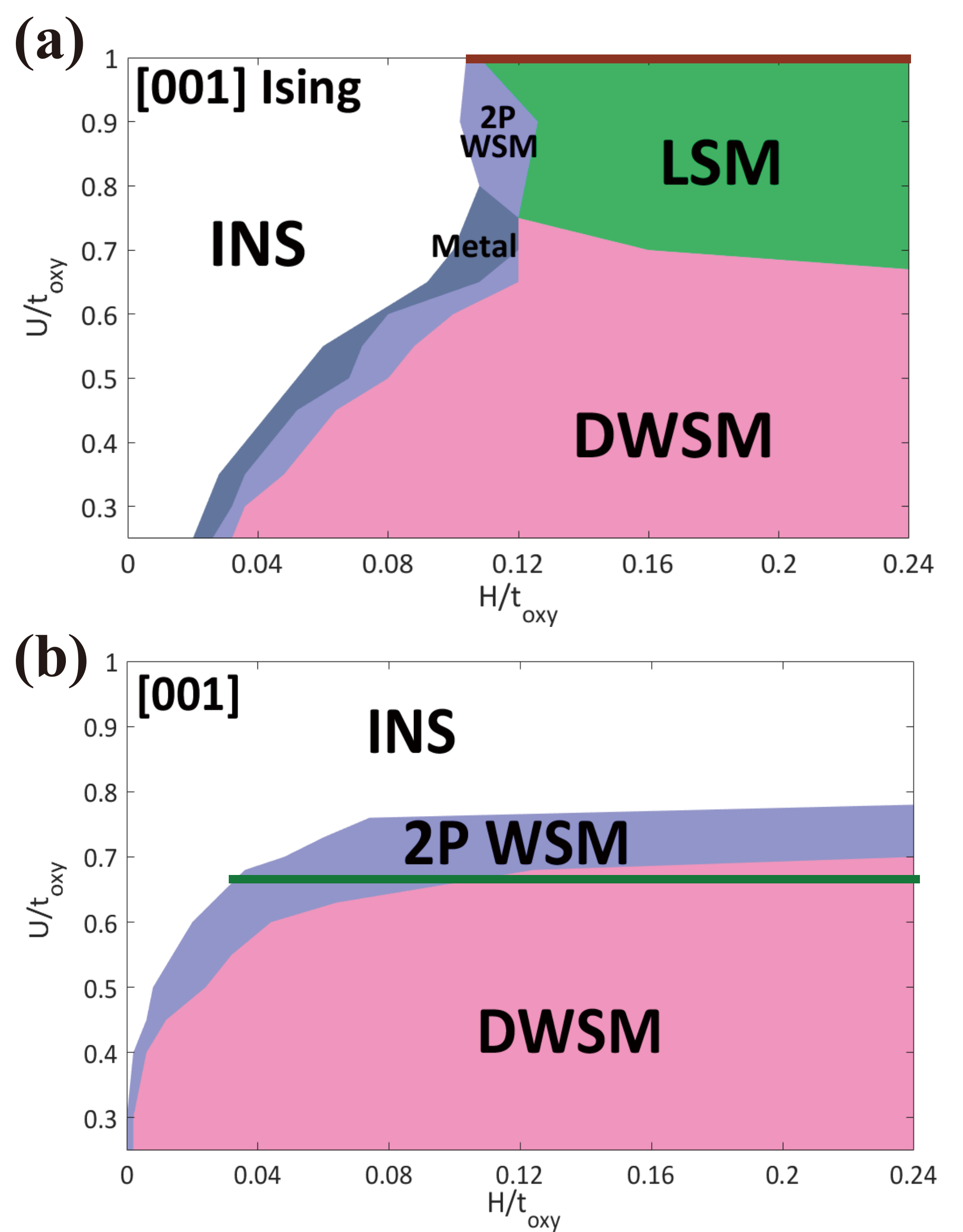}
\caption{
Phase diagrams from self-consistent mean field theory when Ir is treated as an Ising spin (a) or as a Heisenberg spin (b). The horizontal brown (green) solid line in (a) ((b)) corresponds to the brown (green) curve in Fig. 5(a). $\alpha$, $\beta_1$, and $\beta_2$ of the relevant effective Hamiltonian are calculated by using the projection matrix in Eq. (5).
}
\label{fig:7}
\end{figure}
%%%%%%%%%%%%%%%%%%%%%%%%%%%%%%%%%%%%%%%%%%%%%%%

In real materials, the spin modulation pattern under magnetic field depends strongly on the microscopic parameters determining $\vec{B}_{\text{eff,i}}$ in self-consistent calculations.
Fig.~\ref{fig:7} shows two phase diagrams in the $(H,U)$ plane determined by self-consistent mean field theory.
Depending on whether Ir spins are treated as an Ising spin or a Heisenberg spin, we obtain different phase diagrams including distinct topological semimetals.
In both cases, however, the origin of emergent topological semimetals can be understood
based on Fig.~\ref{fig:5}(a). For instance, the mean field Hamiltonian projected along the brown (green) horizontal line in the left (right) figure in Fig.~\ref{fig:7} gives the brown (green) trajectory in Fig.~\ref{fig:5}(a), demonstrating the origin of the relevant topological semimetals. This shows that various emergent topological semimetals can be successfully described by the QBC of the PM coupled to competing magnetic energy scales $\alpha$, $\beta_{1}$, $\beta_{2}$ in the extended Luttinger model.

Up to now, we have considered only Nd, which is a Kramers ion, for the description of $fd$-exchange coupling. However, the role of the non-Kramers ion Pr should be properly taken into account for the application of our theory to (Nd$_{1-x}$Pr$_x$)$_2$Ir$_2$O$_7$ near the QCP. Due to the distinct symmetry properties of Nd and Pr pseudo-spins, the form of $fd$-exchange coupling is also quite different in two cases~\cite{lee2012generic}. For instance, the in-plane components of the pseudospin operators are time-reversal invariant quadrupoles for Pr$^{3+}$ whereas they are time-reversal odd dipole-octupoles for Nd$^{3+}$\cite{huang2014quantum}. As a result, Pr in-plane spin components couple to Ir charge density instead of Ir spin density\cite{lee2013RKKY}. (See \ref{sec:App4}.) However, such a variation in the $fd$-exchange coupling can at most modify the trajectory that the system follows under magnetic field, which can be captured in the variation of $\alpha, \beta_1, \beta_2$ in the extended Luttinger Hamitonian. The global structure of the phase diagram should remain invariant as summarized in Fig.~\ref{fig:5}(a).

\section{Conclusions}

We have shown that magnetic field induced topological semimetals near the QCP can be understood based on the band structure near the $\Gamma$ point. In systems located away from the QCP, however, one need to consider accidental band crossings away from the $\Gamma$ point, which changes the total number of WPs.
For instance, the influence of band crossings at the $L$ point, is shown in \ref{sec:App2}.

Since the presence of QBC near the Fermi level is the key ingredient for the field induced topological semimetals summarized in Fig.~\ref{fig:5}(a), the same idea can be applied to a broad class of materials having a similar low energy band structure, such as HgTe\cite{murakami2004spinhall} or GdPtBi\cite{cano2017chiral}. However, it is worth noting that the non-coplanar magnetic structure of pyrochlore iridates plays a critical role to enlarge the anisotropic Zeeman term in the effective Hamiltonian because it is proportional to the cluster magnetic octupole moment as shown in Eq.~(\ref{eq:5}). Since HgTe is a paramagnet and GdPtBi is a collinear antiferromagnet, the conventional linear Zeeman term should dominate over the Luttinger q-term in both materials, and thus the accessible topological semimetal phases are expected to be more limited.

Since the Pr doping necessarily introduces at least weak disorder effect in the system, although the quality of the pyrochlore iridate samples synthesized recently is reasonably high,
we discuss about the influence of disorder on the phase diagram in Fig.~\ref{fig:5}(a).  
Let us note that because the applied magnetic field lowers the crystalline symmetry, all the topological semimetals shown in Fig.~\ref{fig:5}(a) develop small electron or hole pockets with the nodal points or lines located away from the Fermi level. As it is well known in conventional metals, the weak disorder is an irrelevant perturbation, and thus its influence is negligible. Even if the Weyl points are accidentally located at the Fermi level, weak disorder is still marginally irrelevant according to the recent renormalization group analysis~\cite{isobe2012theory, isobe2013renormalization,yang2014quantum,goswami2011quantum}. Although the disorder effect in a nodal line semimetal is more subtle~\cite{wang2017disorder}, since the gap-closing points of a nodal line generally do not appear simultaneously at the Fermi level and additional small Fermi surfaces from Weyl points are present, we expect that the weak disorder is still irrelevant in the NLS phase as well. Therefore we believe that the physics we have proposed remains valid even in the presence of weak disorder.

We conclude with discussing magnetic fluctuation effects near the QCP\cite{moon2013non,savary2014new}.
Poor screening of Coulomb interaction in the semimetal with QBC is known to induce non-Fermi liquid behavior and unusual magnetic quantum criticality associated with AIAO ordering.
In the presence of magnetic field, however, broken cubic lattice symmetry allows the system to develop electron or hole pockets near Fermi energy $E_{F}$.
In fact, all the topological semimetals shown in Fig.~\ref{fig:5}(a) possess Fermi surface with nodal points or lines located near $E_{F}$. In this case, the magnetic transition of AIAO ordering is described by the conventional Hertz-Millis theory coupled to fermions with Fermi surface. 
To examine the magnetic field induced crossover from non-Fermi liquid physics to conventional Hertz-Millis type behavior and the influence of the bulk topological property on magnetic quantum criticality would be an interesting topic for future study.

% **************************************** Acknowledgement ********************************************** %
\section*{ACKNOWLEDGEMENT}
T. Oh was supported by the Institute for Basic Science in Korea (Grant No. IBS-R009-D1).
H.I. was supported by JSPS KAKENHI Grant
Numbers JP16H06717, JP18H03676, JP18H04222, and
JP26103006, ImPACT Program of Council for Science,
Technology and Innovation (Cabinet office, Government of
Japan), and CREST, JST (Grant No. JPMJCR16F1).
B.-J.Y. was supported by the Institute for Basic Science in Korea (Grant No. IBS-R009-D1) and Basic Science Research Program through the National Research Foundation of Korea (NRF) (Grant No. 0426-20170012, No.0426-20180011), the POSCO Science Fellowship of POSCO TJ Park Foundation (No.0426-20180002), and the U.S. Army Research Office under Grant Number W911NF-18-1-0137.
We thank N. Nagaosa for useful discussion.

\renewcommand{\thesection}{APPENDIX~\Alph{section}}
\renewcommand{\thesubsection}{\arabic{subsection}}
\renewcommand{\thefigure}{S\arabic{figure}}
\renewcommand{\thetable}{S\arabic{table}}
\renewcommand{\theequation}{S\arabic{equation}}
\setcounter{section}{0}
\setcounter{equation}{0}
\setcounter{table}{0}
\setcounter{figure}{0}

\section{Effective Theory at $\Gamma$ Point \label{sec:App1}}

\subsection{Symmetry of Pyrochlore Iridates}

Pyrochlore iridate $R_2 Ir_2 O_7$ (R-227) comprises two intertwined pyrochlore lattices of $R$ (rare-earth) and $Ir$ ions. An octahedron with oxygen ions surrounds each $Ir$ ions. Each pyrochlore lattice is composed of linked tetrahedra, in which two adjacent tetrahedra are inversion-symmetric about the linked point. Fig. 1(a) shows the structure of pyrochlore iridates. A tetrahedron is the unit cell of pyrochlore lattice. The structure of pyrochlore iridates is depicted in Fig.~\ref{fig:1}(a), 

The point group of pyrochlore iridates is $T_d$ (tetrahedron), which contains 5 equivalent classes: identity ($\mathbb{I}$), 3-fold rotations ($C_3$), twofold rotations ($C_2$), diagonal mirrors ($\sigma_d$), $\pi$/2 rotations followed by mirrors ($S_4$). Including spin-orbit coupling (SOC) in the system, we should utilize $T_d$ double group in the argument. $T_d$ double group has 8 equivalent classes, including identity ($\bar{\mathbb{I}}$), 3-fold rotations($\bar C_3$), and $\pi/2$ rotations followed by mirrors($\bar S_4$) after $2\pi$-rotation. Accordingly, the number of $T_d$ double group representations is 8. Moreover, $\mathbb{P}\equiv\{P|T_{1/4,1/4,1/4}\}$ (space inversion and half-translation), $T$ (time-reversal), and $T_r$ (FCC lattice translation) symmetries are preserved. The space group of pyrochlore iridates is $fd\bar{3}m$. Since we argue in momentum space, $\mathbb{P}\equiv P$ regardless of eigenvalues.

%******************************* Luttinger Hamiltonian ******************************%
\subsection{Luttinger Hamiltonian \label{sec:symmetrygroup}}

We begin with quadratic band crossing in the paramagnetic semimetal phase of $Pr$-227 \cite{EGM2015NaCom}. Since the magnetic ordering simultaneously occurs with metal-insulator transition, we infer that magnetic ordering is the crucial source of band manipulation. Thus, we can assume, in general, quadratic band crossing appears for the paramagnetic semimetal phase of pyrochlore iridates. 

With $T$ and $P$ symmetry in the system, one needs at least $4 \times 4$ Hermitian matrices by Kramers degeneracy of each band. According to group theory, we should use $4 \times 4$ Hermitian matrices, since the largest dimension among the irreducible representations (irreps) of $T_d$ double group is 4 ($\Gamma_8$ representation). 

We can bulid effective Hamiltonian by gathering the anti-commuting matrices since such Hamiltonian gives only two distinct energy bands. The number of basis of the space of $4 \times 4$ Hermitian matrices is 16, but only 5 of them are anti-commuting. Therefore, the effective Hamiltonian of quadratic band crossing is
\begin{align}
\mathcal H_0 (\vec{k}) = \epsilon_0(\vec{k}) + \sum_{i=1}^{5} d_i(\vec{k})\Gamma_i
\end{align}
where $\epsilon_0(\vec{k}) = k^2/2m$, and $\Gamma_i$ are 5 anti-commuting $4\times4$ matrices, $\{\Gamma_i, \Gamma_j\} = 2\delta_{ij}$. The algebra is called $SO(5)$ Clifford algebra. Explicitly, 
\begin{align}
\Gamma_1 =& \f{1}{\sqrt{3}}\{J_y,J_z\} = \ \tau_z\sigma_y,\ \Gamma_2 = \f{1}{\sqrt{3}}\{J_z,J_x\} =\tau_z\sigma_x\n
\Gamma_3 =& \f{1}{\sqrt{3}}\{J_x,J_y\} = \tau_y,\Gamma_4 = \f{1}{\sqrt{3}}(J_x^2-J_y^2) = \tau_x,\n
\Gamma_5 =& J_z^2-\f{5}{4}=\tau_z\sigma_z,
\end{align}
where $\tau_i$, $\sigma_i$ are Pauli matrices, and $J_i$ are spin-$3/2$ matrices. Also, the coefficients are defined as
\begin{align}
d_1 =& -\sqrt{3}a k_yk_z,\ d_2 = -\sqrt{3}ak_zk_x,\ d_3 = -\sqrt{3}ak_xk_y,\n
d_4 =& -\frac{\sqrt{3}}{2}b(k_x^2-k_y^2),\ d_5 = -\frac{1}{2}b(2k_z^2-k_x^2-k_y^2),
\end{align}
where $a, b$ are arbitrary constants. 

The Hamiltonian is called Luttinger Hamiltonian\cite{luttinger1956quantum}. Since we are only interested in the band crossings, we assume particle-hole symmetry and isotropy, for convenience. That is, we ignored the term $\epsilon_0(\vec{k})$ and let the coefficient $a=b=1$,  Furthermore, we concentrate on the band crossing between two middle bands, since we will assume the half-filling in the lattice model.

%******************************* AIAO order of magnetic moment ******************************%
\subsection{AIAO order parameter}

%Since the magnetic order in the lattice changes the band structure, the magnetic order in the effective theory.
As neutron scattering experiment turned out, rare-earth or $Ir$ moments in pyrochlore iridates form all-in-all-out (AIAO) order \cite{tomiyasu2012emergence}, in which every magnetic moment points either to or from the center of the tetrahedron (Fig. \ref{fig:1}(b)). Accordingly, we should primarily include AIAO order parameter in the theory. 
A pyrochlore lattice with AIAO order breaks $T$, $\sigma_d$, and $S_4$ symmetry, but preserves the combinations, $\sigma_d T$ and $S_4 T$.

AIAO order parameter transforms as the $\Gamma_2$ representation of $T_d$ double group. Hence, we should add
\begin{align}
\mathcal H_{AIAO} = - \alpha \Gamma_{45},
\end{align}
where $\Gamma_{ab} = [\Gamma_{a}, \Gamma_{b}]/2i$, and $\alpha$ is AIAO order parameter.

In presence of AIAO order only, the effective Hamiltonian is 
\begin{align}
\mathcal H_{eff,1} = \mathcal H_0 + \mathcal H_{AIAO}. \label{eq:effective}
\end{align}
The eigenenergy is
\begin{align}
E_{\eta,\zeta} = \eta\sqrt{k^4 + \alpha^2 + 2\alpha\zeta \sqrt{d_1^2+d_2^2+d_3^2}}
\end{align}
where $\eta, \zeta = \pm1$. $(\eta,\zeta\frac{|\alpha|}{\alpha}) = (+1,-1)$ and $(-1,-1)$ cross at eight Weyl points, $\vec{k} = \sqrt{|\alpha|/3}\ (\pm1,\pm1,\pm1)$. Weyl points stick on 3-fold rotation invariant($[H,C_3] = 0$) axes, $[111],[1\bar1\bar1],[\bar11\bar1]$, and $[\bar1\bar11]$, for any $\alpha\neq0$. According to the condition $C_3^3=-1$, 3-fold rotation operator can have three distinct eigenvalues, $e^{\pm i\pi/3}$ and $-1$, and two crossing bands have different eigenvalues among them. For example, for the band crossing along [111] direction, the eigenvalues of crossing bands are $e^{-i\pi/3}$ and $-1$, respectively.

\subsection{Effective field}

%Beyond AIAO order parameter, there are 11 other classfied CMMMs. Such symmetrized CMMMs can appear when magnetic moment configuration changes. For example, magnetic field is a source to change the magnetic moment. Providing the symmetry under magnetic field and AIAO order, we can reduce the number of symmetrized CMMMs. We regard two cases that $[001]$ or $[111]$ direction of magnetic field is applied.

Before arguing the topological phases under effective field, we must note the remaining symmetries for each direction of field.

If magnetic field is applied in $[001]$ direction without any magnetic order in the pyrochlore iridates, the symmetry operations are identity $I$, twofold rotation $C_{2z}$, twofold rotation followed by time reversal $C_{2x}T, C_{2y}T$, and the mirror symmetry about the plane including $[001]$ followed by time-reversal $2\sigma_{d,001}T$, $\pi/2$ rotation followed by the mirror $S_{4z}$, and inversion ($P$). A combined symmetry, $k_z=0$ plane mirror $\mathcal{M}_z = C_{2z}P$ also exist. $\mathcal{M}_z$ acts as mirror symmetry only in momentum space, since $P$ is inversion with half-translation in real space. If we apply $[001]$ direction field with AIAO order of magnetic moment, there are only $I$, $C_{2z}$, $2\sigma_{d,001}T$, $P$, and $\mathcal{M}_z$.

If magnetic field is applied in $[111]$ direction without any magnetic order in the pyrochlore iridates, the symmetry operations are identity $I$, 3-fold rotation around [111] line $C_{3,111}$, mirrors through the plane including $[111]$ followed by time-reversal $3\sigma_{d,111}T$, and inversion $P$. If we apply $[111]$ direction field with AIAO order of magnetic moment, still $I$, $C_{3,111}$, $3\sigma_{d,111}T$, and $P$ are preserved. 

The magnetic field transforms as $\Gamma_4$ representation of $T_d$ double group, so the following terms are allowed.
\begin{align}
\mathcal H_{B} =& % - (\beta_1^h \vec{H} + \beta_1^m \vec{M} + \beta_1^a \V A) \cdot \vec{J_1} - (\beta_2^h \vec{H} + \beta_2^m \vec{M} + \beta_2^a \V A) \cdot \vec{J_3} \n 
-\beta_1 \vec{B} \cdot \vec{J_1} - \beta_2 \vec{B} \cdot \vec{J_3} \label{eq:Zeeman},
\end{align}
where $\V B(\V H, \V M, ...)$ is the effective magnetic field, which is the function of magnetic field $\V H$, and magnetization $\V M$, and other order parameters which transforms as same as magnetic field and magnetization.
\begin{align}
\mathcal H_{eff,2} = \mathcal H_0 + \mathcal H_B. \label{eq:effective2}
\end{align}
By diagonalizing $\mathcal H_{eff,2}$, we can observe topological phases when AIAO order parameter is trivial. 

Although $\mathcal H_{eff,2}$ is too complicated to obtain the energy spectrum in an analytic way, we can acquire the energy spectrum along high-symmetry lines and on the mirror planes. Let us consider [001] effective field first. Then, Eq.~\ref{eq:Zeeman} becomes
\begin{align}
\mathcal H_{B,001} = - B(\cos \theta J_z + \sin\theta J_z^3), \label{eq:001B}
\end{align}
where $\beta_1 = \cos\theta, \beta_2=\sin\theta$. $\theta$ is the variable that controls the relative magnitude of Zeeman and Luttinger q-term. Since there is $C_{2z}$ and $\mathcal{M}_z = C_{2z}P$ symmetry, we investigate along $k_z$-axis and $k_z=0$ plane.

Along $k_z$-axis, the Hamiltonian becomes
\begin{align}
\mathcal H_{001} = d_5(k_z) \Gamma_5 + \mathcal H_{B,001}.
\end{align}
because $d_{1,2,3,4}=0$ when $k_x=k_y=0$. Since the Hamiltonian is already diagonalized on the basis of $J_z = \pm 3/2, \pm 1/2$, the energy spectrum is just
\begin{align}
&E_{3/2} = k_z^2 - B(\frac{3}{2}\cos\theta + \frac{27}{8}\sin\theta) \n
&E_{-3/2} = k_z^2 + B(\frac{3}{2}\cos\theta + \frac{27}{8}\sin\theta) \n
&E_{1/2} = -k_z^2 - B(\frac{1}{2}\cos\theta + \frac{1}{8}\sin\theta) \n
&E_{-1/2} = -k_z^2 + B(\frac{1}{2}\cos\theta + \frac{1}{8}\sin\theta).
\end{align}
According to the energy spectrum at $\Gamma$, the band crossings of two middle bands will change as varying $\theta$.

Defining $a \equiv \frac{3}{2}\cos\theta + \frac{27}{8}\sin\theta$ and $b \equiv \frac{1}{2}\cos\theta + \frac{1}{8}\sin\theta$, we can divide into 4 cases, where $a$ and $b$ is either positive or negative, respectively. Note that $\mu_i$ are Pauli matrices, $B$ is positive, and $\theta$ ranges from $0$ to $2\pi$.

\begin{enumerate}
\item$a>0,b>0$

If $\theta < \pi-\arctan 4 = \theta_{b1}$ or $\theta > 2\pi-\arctan\frac{4}{9} = \theta_{b4}$, then $a>0, b>0$. $E_{3/2}$ and $E_{-1/2}$ are two middle bands, and cross at two points. Concentrating on two crossing bands, we induce
\begin{align}
\mathcal H_{1,2\times2} = &\frac{-a+b}{2}\mu_0 + \frac{-a-b}{2}\mu_z + d_3(\vec k) \mu_y \n
&+ d_4(\vec k) \mu_x + d_5(\vec k) \mu_z.
\end{align}
A pair of double Weyl points emerge according to the d-wave nature of $d_3$ and $d_4$.

\item$a<0,b<0$

If $\pi-\arctan \frac{4}{9} = \theta_{b2} < \theta < 2\pi - \arctan 4 = \theta_{b3}$, then $a<0, b<0$. $E_{-3/2}$ and $E_{1/2}$ are two middle bands, and cross at two points. The two-band projected theory is
\begin{align}
\mathcal H_{2,2\times2} = &\frac{a-b}{2}\mu_0 + \frac{-a-b}{2}\mu_z + d_3(\vec k) \mu_y \n&+ d_4(\vec k) \mu_x - d_5(\vec k) \mu_z.
\end{align}
For the same reason as the first case, a couple of double Weyl points appear.

\item $a>0, b<0$

When $\theta_{b1} < \theta < \theta_{b2}$, we have $a>0, b<0$. $E_{3/2}$ and $E_{1/2}$ are two middle bands, and cross at two points. Two crossing bands are written as
\begin{align}
\mathcal H_{3,2\times2}= &\frac{-a-b}{2}\mu_0 + \frac{-a+b}{2}\mu_z + d_1(\vec k) \mu_y \n&+ d_2(\vec k) \mu_x + d_5(\vec k) \mu_z.
\end{align}
In this case, $d_1$ and $d_2$ have p-wave nature, so a pair of single Weyl points emerge.

\item $a<0,b>0$

If $\theta_{b3} < \theta < \theta_{b4}$, we have $a<0, b>0$. $E_{-3/2}$ and $E_{-1/2}$ are two middle bands, and cross at two points. Projecting on two crossing bands, we have
\begin{align}
\mathcal H_{4,2\times2}= &\frac{a+b}{2}\mu_0 + \frac{-a+b}{2}\mu_z - d_1(\vec k) \mu_y \n& - d_2(\vec k) \mu_x - d_5(\vec k) \mu_z.
\end{align}
Likewise, there are a couple of single Weyl points.
\end{enumerate}

To sum up, along $k_z$-axis, $\theta$ determines the emergence of either a pair of double Weyl or single Weyl points. All of them stick at $k_z$-axis by twofold rotation symmetry $C_{2z}$, whose possible eigenvalues are $\pm i$. The eigenvalues of twofold rotation operator of crossing bands are equal for double Weyl, but opposite for single Weyl points. These points are topologically protected $-$ even though twofold rotation symmetry is broken, Weyl points are not annihilated immediately.

Meanwhile, on $k_z=0$ plane, Hamiltonian becomes
\begin{align}
\mathcal H_{k_z=0}(\vec k) =& d_3(\V k) \Gamma_3 + d_4(\V k) \Gamma_4 + d_5(\V k) \Gamma_5 \n& + \mathcal H_{B,001},
\end{align}
where $\vec k = (k_x,k_y,0)$. The plane is invariant under $\mathcal{M}_z=C_{2z}P$; that is, $\mathcal{M}_z^{\dagger} H_{k_z=0}(k_x,k_y,0)\mathcal{M}_z = H_{k_z=0}(-k_x,-k_y,0) = H_{k_z=0}(k_x,k_y,0)$. Diagonalizing the matrix, then we get the energy spectrum
\begin{align}
E_{\eta,\zeta} = &\eta\sqrt{d_3^2+d_4^2+[d_5-\zeta B(\cos\theta+\frac{7}{4}\sin\theta)]^2}\n&-\zeta B(\frac{1}{2}\cos\theta + \frac{13}{8}\sin\theta),
\end{align}
where $\eta, \zeta = \pm 1$. $-i\zeta$ is an eigenvalue of $\mathcal{M}_z$. In order to detect band crossings, let us define $c \equiv (\frac{1}{2}\cos\theta + \frac{13}{8}\sin\theta)$ and $d \equiv |\cos\theta+\frac{7}{4}\sin\theta|$. Since $B$ and $d$ are positive-definite, we divide the case according to the sign of $c$.

To observe crossing points, we should consider two aspects, $\mathcal{M}_z$ eigenvalue of each band and energy level at $\Gamma$. Two crossing bands should have different eigenvalues of $\mathcal{M}_z$, either $-i$ or $+i$. In addition, the lowest/highest part of the crossing band which appears at $\Gamma$ should be in the negative/positive energy level, since energy increases/decreases monotonically when departing from $\Gamma$. 
\begin{enumerate}
\item When $c>0$, there are the crossings between $E_{1,1}$ and $E_{-1,-1}$ emerge only if $c-d >0$, where $\mp(c-d)$ is the $\Gamma$ point energy of each band. $\mathcal{M}_z$ eigenvalue of $E_{1,1}$ and $E_{-1,-1}$ is $-i$ and $+i$, respectively. Surprisingly, the range of $\theta$ satisfying $c-d>0$ is $\theta_{b1} < \theta < \theta_{b2}$, which is the range of $a>0, b<0$. 

\item If $c<0$, there are the crossings between $E_{1,-1}$ and $E_{-1,1}$ when $c+d < 0$, where $\pm(c+d)$ is the $\Gamma$ point energy of each band. Likewise, $\mathcal{M}_z$ eigenvalue of $E_{1,-1}$ and $E_{-1,1}$ is $+i$ and $-i$, respectively. The range of $\theta$ satisfying $c+d<0$ is $\theta_{b3} < \theta < \theta_{b4}$, which is consistent with the range of $a<0, b>0$. 
\end{enumerate}
We infer that the crossings on $k_z=0$ plane coexist with a pair of single Weyl points on $k_z$-axis, and are protected by $\mathcal{M}_z$ symmetry. 

$\mathcal{M}_z$ symmetry makes a line node form crossing on $k_z=0$ plane. For example, for $c>0$ case, the crossing occurs when
\begin{align}
k_x^2+k_y^2 = 2\frac{\sqrt{c^2(4c-d)(c-d))}}{d^2-4c^2}
\end{align}
This is nothing but a circle. The term inside the square root can only be positive only if $c-d>0$. With a similar argument, we also have a line node for $c<0$ case.

\begin{figure}
\centering
\includegraphics[width=\columnwidth]{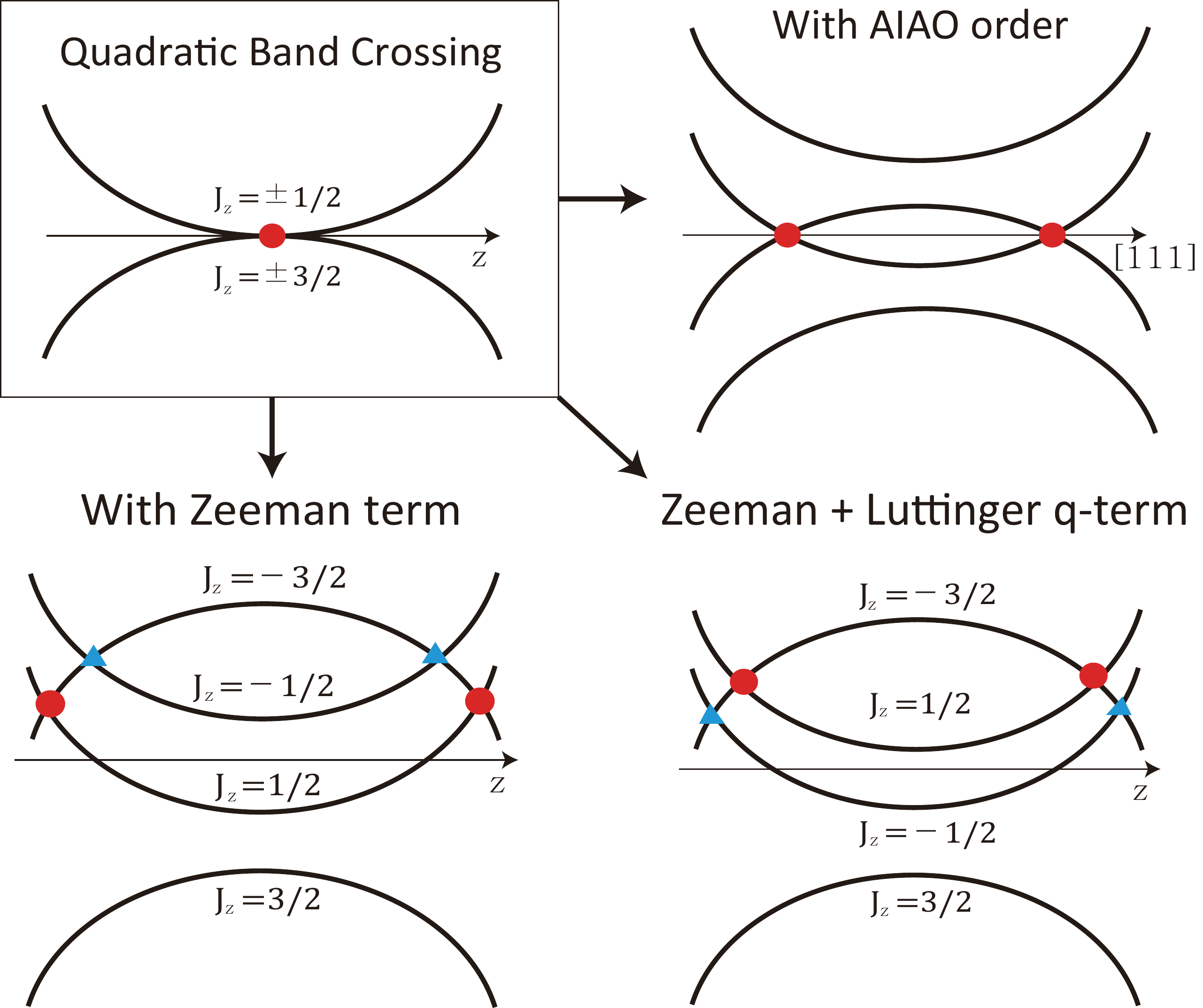}
\caption{The role of AIAO order parameter, Zeeman term, and Luttinger q-term in quadratic band crossing. Only with AIAO or Zeeman term, we find no additional level crossing between 4 degenerate eigenstates at $\Gamma$ point. However, with both Zeeman and Luttinger q-term, additional level crossing can occur.}
\label{fig:mechanism}
\end{figure}

In short, with $[001]$ effective field, we observe two available crossings: \textit{i)} A pair of double Weyl points along the $k_z$-axis (Double Weyl semimetal, DWSM), \textit{ii)} A pair of single Weyl points along $k_z$-axis with a line node on $k_z=0$ plane (Line node semimetal, LSM). Crossing points at $k_z$-axis is topologically protected, while a line node on $k_z=0$ plane is protected by $\mathcal{M}_z$ symmetry. With only Zeeman term, we find no additional level crossing between 4 degenerate eigenstates, while in the presence of Zeeman and Luttinger q-term, additional level crossing can occur. (See in Fig.~\ref{fig:3} and \ref{fig:mechanism})

%111 direction

Next, with $[111]$ effective field, the Hamiltonian is
\begin{align}
\mathcal H_{B,111} = &- \frac{B}{\sqrt{3}}[\cos \theta (J_x+J_y+J_z) \n&+ \sin\theta (J_x^3+J_y^3+J_z^3)]. \label{eq:111}
\end{align}
Even on the high-symmetry line $[111]$, notwithstanding, it is too complicated to acquire the energy spectrum analytically. The Hamiltonian cannot be diagonalized under the basis that $J_{111} = \pm \frac{3}{2}, \pm \frac{1}{2}$, since $(J_x+J_y+J_z)$ and $(J_x^3+J_y^3+J_z^3)$ are not commute. In spite of the complexity, we can still investigate $\Gamma$ point energy spectrum to acquire the nature of crossing points between two middle bands on [111] line, where 3-fold rotational symmetry is preserved. 

\begin{figure}
\centering
\includegraphics[width=\columnwidth]{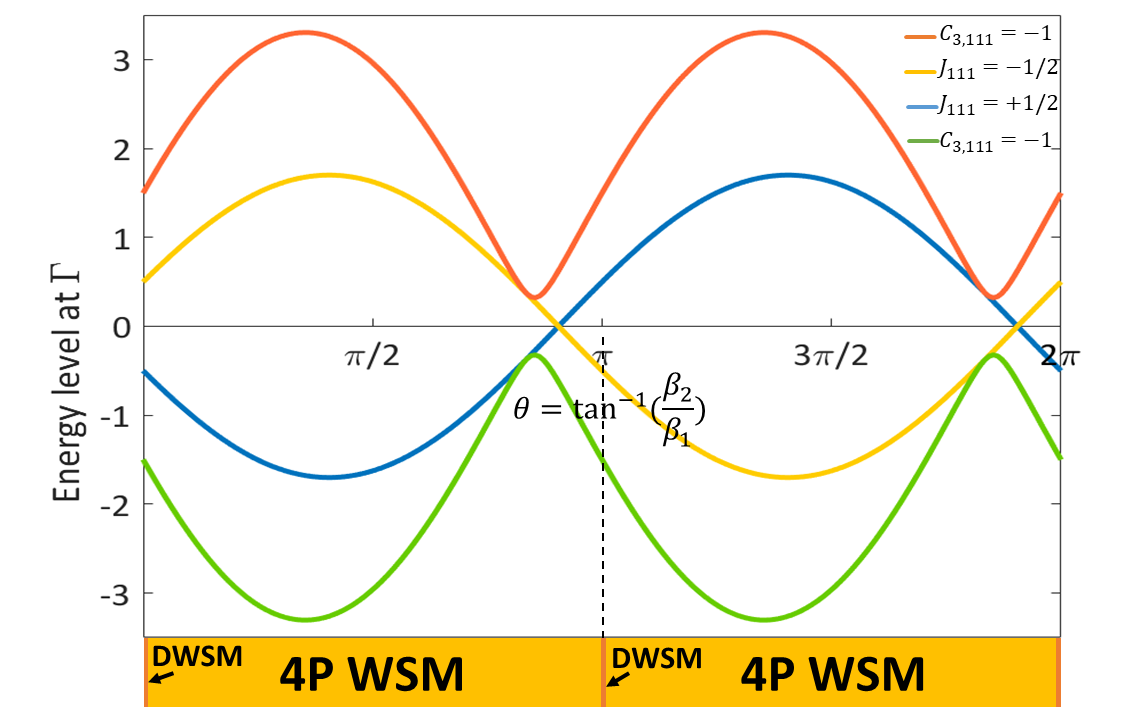}
\caption{The energy spectrum at $\Gamma$ point as a function of $\theta=\tan^{-1}\B_2/\B_1$ is drawn for $[111]$ field. We suppose the angular momentum unit to be 1, and apply the unit effective Zeeman field like in Fig.~\ref{fig:4}. The Hamiltonian is diagonalizable with $C_{3,111}$ eigenstates instead of $J_{[111]}$ eigenstates. As the top and bottom energy level share $C_{3,111}$ eigenvalue, they cannot cross as $\theta$ changes, unlike $[001]$ field.}
\label{fig:111levels}
\end{figure}

In Fig.~\ref{fig:111levels}, four energy levels at $\Gamma$ point has been drawn against $\theta$. At any $\theta$, we can label the energy level at $\Gamma$ with eigenvalues of $C_{3,111}=e^{i\frac{2\pi}{3\sqrt{3}}(S_x+S_y+S_z)}$, 3-fold rotation around $[111]$ line. Comparing $C_{3,111}$ basis with the $J_{[111]}$ eigenstates, we confirm that $J_{[111]}=\pm 1/2$ corresponds to two middle energy levels for every $\theta$, and the other two energy levels are the linear combination of $J_{[111]}=\pm 3/2$. From $C_{3,111}^3 = -1$, we observe that $J_{[111]}=\pm 1/2$ have the eigenvalue of $e^{\mp i\pi/3}$, and $J_{[111]}=\pm3/2$ states have eigenvalue $-1$. Since the number of energy band is larger than the number of possible eigenvalues of $C_{3,111}$, it is natural to have two energy levels whose eigenvalues of $C_{3,111}$ are identical. Furthermore, since $J_{[111]}=\pm 3/2$ have the same eigenvalue of $C_{3,111}$, the hybridization of $J_{[111]}=\pm3/2$ is inevitable.

%%%%%%%%%%%%%% Gamma Energy %%%%%%%%%%%%%%%%%

We label the energy band as $E_{-1,top}, E_{-1,bot}, E_{\pm 1/2}$, whose energy level at $\Gamma$ is in Fig.~\ref{fig:111levels}. We consider $0 \leq \theta < \pi$ range primarily, since $\pi \leq \theta < 2\pi$ will be similar. The band crossing occurs between $E_{-1,top}$ and $E_{1/2}$ for $\theta < \pi-\arctan\frac{4}{13} \sim 2.84309 = \theta_{c2}$, and between $E_{-1,top}$ and $E_{-1/2}$ otherwise. Meanwhile, $E_{-1,top}$ band changes its component from $J_{[111]}={-3/2}$ to ${3/2}$, when $\theta$ varies from $0$ to $\pi$. Especially, $J_{[111]}=\pm 3/2$ states are significantly mixed near $\theta_{c1}=\pi-\frac{1}{2}\arctan(\frac{184}{139}) \sim 2.67968$. The band crossings between $E_{-1,top}$ and $E_{1/2}$ are double Weyl nodes only if $\theta=0$, since $J_{[111]}=\pm 3/2$ becomes the eigenstates of Hamiltonian. Otherwise, each double Weyl node is broken into 4 single Weyl nodes. According to the conservation of topological charge, one of 4 single Weyl node has the opposite topological charge of the rest of single Weyl nodes.

To sum up, band crossings along $[111]$ direction are double Weyl points at $\theta=0$ where $J_{[111]}=-\frac{3}{2}, \frac{1}{2}$ states cross, and immediately breaks into single Weyl points as $\theta$ increases because of the hybridization of $J_{[111]}=\pm 3/2$ states. In fact, observing crossing points numerically, one can locate 4 pairs of single Weyl points including a pair on $[111]$ axis. Similarly, band crossings become double Weyl points at $\theta=\pi$, because $J_{[111]}=\frac{3}{2}, -\frac{1}{2}$ cross.

So far, we suggest a various topological phases of pyrochlore iridates only under effective field, by virtue of the interplay between Zeeman and Luttinger q-term. For $[001]$ direction, Double Weyl semimetal (DWSM) and a line-node semimetal (LSM) emerge. For $[111]$ direction, DWSM and 4-pair Weyl semimetal (4P WSM) appear. 

%******************************* AIAO and Effective Field ******************************%
\subsection{AIAO and effective field}

In this section, we take both AIAO order parameter and effective field into account simultaneously. 

Given large AIAO order parameter and weak effective field strength, we draw trajectories of the crossing points through the perturbation theory near each Weyl points. In addition, we investigate the emergence of crossing points between two middle bands by the perturbation theory near $\Gamma$, and establish the phase diagram with two variables: $\theta$, controling the ratio between Zeeman and Luttinger q-term, $\phi$, controling the ratio between AIAO order parameter and effective field strength ($\tan\phi = B/\alpha$).

%%%%%%%%%%%%001 direction

\subsubsection{[001] direction \label{[001]}}

Let us begin from $[001]$ direction. There are 8 Weyl points if AIAO order exist, and all of them stick on 3-fold rotation axis. Since $[001]$ magnetic field breaks all of 3-fold rotation symmetries, every Weyl points will move away from the rotation axes. Given the mirror symmetry $2\sigma_{d,001}$ and the topological nature, Weyl points will travel on the mirror plane whose normal vector is either $[110]$ or $[1\bar10]$. If Weyl points travel out of the plane, each Weyl nodes should divide into two by mirror symmetry, then Nielsen-Ninomiya Theorem is violated.

According to the symmetries, we can divide 8 Weyl fermions into 2 classes: Class 1, 4 Weyl points included in the mirror plane with $[1\bar10]$ normal vector, and Class 2, other 4 Weyl points included in the mirror plane with $[110]$ normal vectors.

If we choose one of Class 1 Weyl points at $\vec k_{C1,001} = \sqrt{\frac{\alpha}{3}}(1,1,1)$, the Hamiltonian near the point $\vec k = \vec k_{C1,001} + \vec q$  becomes 
\begin{align}
\mathcal H_{C1}^{001} = \mathcal H_{0,C1}^{001}+ \mathcal H_{mom,C1}^{001},
\end{align}
where
\begin{align}
\mathcal H_{0,C1}^{001} =& -\frac{\alpha}{\sqrt{3}}(\Gamma_1 + \Gamma_2 + \Gamma_3) - \alpha\Gamma_{45}. \label{eq:H0C1}\\
\mathcal H_{mom,C1}^{001} =&  -\sqrt{\alpha}[(q_y+q_z)\Gamma_1 + (q_z+q_x)\Gamma_2\n&+ (q_x+q_y)\Gamma_3 + (q_x-q_y)\Gamma_4 \n&+ \frac{1}{\sqrt{3}}(2q_z-q_x-q_y)\Gamma_5], \label{eqmom1}
\end{align}
up to the first order of $\V q =(q_x,q_y,q_z)$. In addition, we apply the second-order degenerate perturbation theory on magnetic field Hamiltonian (Eq. \ref{eq:001B}). We denote $\gamma_1=B\cos\theta$ and $\gamma_2=B\sin\theta$. Concentrating on two crossing bands, we obtain the following effective model.
\begin{align}
\mathcal H_{proj,C1}^{001} = A_0\sigma_0 + A_1 \sigma_x + A_2 \sigma_y + A_3 \sigma_z,
\end{align}
where
\begin{align}
A_0 =& \frac{\sqrt{3}}{6}\gamma_1 + \frac{13\sqrt{3}}{24}\gamma_2 \nonumber \\
A_1 =& -\sqrt{\frac{\alpha}{6}}(-(2-\sqrt{3})q_x-(2+\sqrt{3})q_y-2q_z) - \frac{\gamma_1^2}{4\sqrt{2}\alpha} \n&- \frac{9\gamma_2^2}{64\sqrt{2}\alpha} \n
A_2 = & -\sqrt{\frac{\alpha}{6}}((2+\sqrt{3})q_x+(2-\sqrt{3})q_y+2q_z) + \frac{\gamma_1^2}{4\sqrt{2}\alpha} \n&+ \frac{9\gamma_2^2}{64\sqrt{2}\alpha} \n
A_3 = & \sqrt{\frac{\alpha}{3}}(q_x+q_y-2q_z) + \frac{\sqrt{3}\gamma_1}{3} + \frac{7\sqrt{3}\gamma_2}{120}. \nonumber
\end{align}
Weyl points will exist when $A_1=A_2=A_3=0$. The solutions are
\begin{align}
q_x =&~ q_y = \frac{-16\alpha(4\gamma_1+7\gamma_2)+\sqrt{3}(16\gamma_1^2+9\gamma_2^2)}{384 \alpha^{3/2}},\n q_z =&~ \frac{32\alpha(4\gamma_1+7\gamma_2)+\sqrt{3}(16\gamma_1^2+9\gamma_2^2)}{384\alpha^{3/2}}.
\end{align}
The rotation symmetry $C_{2z}$ and inversion $P$ determine the trajectory of other 3 Class 1 Weyl points.

\begin{figure}
\centering
\includegraphics[width=\columnwidth]{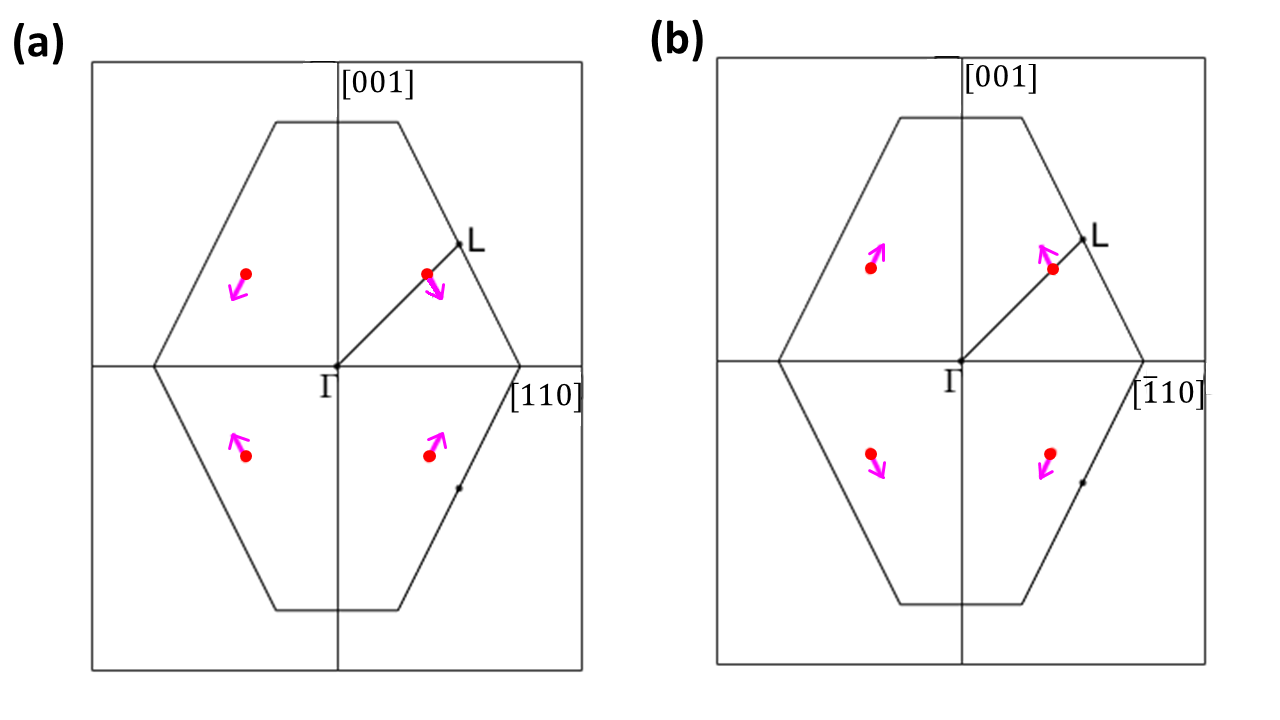}
\caption{(a) Under $[001]$ magnetic field, the trajectories of Class 1 Weyl points on the mirror $[1\bar10]$ plane, and (b) Class 2 Weyl points are drawn on the mirror $[110]$ plane with colored lines. ($\gamma_1=\gamma_2$). The black solid line is a virtual boundary of Brillouin Zone to present the trajectories of Weyl points in an obvious way.}
\label{fig:Traj001}
\end{figure}

Meanwhile, at one of Class 2 Weyl points $\vec k_{C2,001} = \sqrt{\frac{\alpha}{3}}(-1,1,1)$, the Hamiltonian at $\vec k = \vec k_{C2,001}+\vec q$ is
\begin{align}
\mathcal H_{C2}^{001} = \mathcal H_{0,C2}^{001}+\mathcal H_{mom,C2}^{001}.
\end{align}
where
\begin{align}
\mathcal H_{0,C2}^{001} =& \frac{\alpha}{\sqrt{3}}[-\Gamma_1+\Gamma_2+\Gamma_3]-\alpha\Gamma_{45} \\
\mathcal H_{mom,C2}^{001} =& -\sqrt{\alpha}[(q_y+q_z)\Gamma_1 + (q_x-q_z)\Gamma_2 \n&+ (q_x-q_y)\Gamma_3 - (q_x+q_y)\Gamma_4 \n&+ \frac{1}{\sqrt{3}}(2q_z+q_x-q_y)\Gamma_5],
\end{align}
up to the first order of $\V q$. By the same procedure as Class 1, we obtain the following Hamiltonian.
\begin{align}
\mathcal H_{proj,C2}^{001} = B_0 \sigma_0 + B_1 \sigma_x + B_2\sigma_y + B_3 \sigma_3,
\end{align}
where
\begin{align}
B_0 =& - \frac{\sqrt{3}\gamma_1}{6} - \frac{13\sqrt{3}\gamma_2}{24} \nonumber \\
B_1 =& -\sqrt{\frac{\alpha}{6}}(-(2+\sqrt{3})q_x+(2-\sqrt{3})q_y+2q_z) \n&+ (\frac{\gamma_1^2}{\alpha}\frac{1}{4\sqrt{2}}+\frac{\gamma_2^2}{\alpha}\frac{9}{64\sqrt{2}}) \nonumber \\
B_2 =&-\sqrt{\frac{\alpha}{6}}(-(2-\sqrt{3})q_x+(2+\sqrt{3})q_y+2q_z) \n&+ (\frac{\gamma_1^2}{\alpha}\frac{1}{4\sqrt{2}}+\frac{\gamma_2^2}{\alpha}\frac{9}{64\sqrt{2}}) \nonumber \\
B_3 =& -\sqrt{\frac{\alpha}{3}}(q_x-q_y+2q_z) - (\gamma_1\frac{\sqrt{3}}{3} + \gamma_2\frac{7\sqrt{3}}{12}). \nonumber 
\end{align}
Therefore, Weyl points are at
\begin{align}
q_x =&~ -q_y = -\frac{16\alpha(4\gamma_1+7\gamma_2)+\sqrt{3}(16\gamma_1^2+9\gamma_2^2)}{384\alpha^{3/2}},\n
q_z =&~ \frac{-32\alpha(4\gamma_1+7\gamma_2)+\sqrt{3}(16\gamma_1^2+9\gamma_2^2)}{384a^{3/2}}.
\end{align}
Other 3 Class 2 Weyl points are determined by 3-fold rotation and inversion symmetry. The result implies that Weyl points can only move on the mirror plane $[1\bar10]$ or $[110]$, and the direction of trajectory of each Class is distinct. The trajectories are drawn in Fig. \ref{fig:Traj001}.

%%%%%%%%%%%%% Trajectory 001 %%%%%%%%%%%%%

From now on, we introduce two variables $\theta$ and $\phi$. The effective theory at $\Gamma$ point with AIAO order and $[001]$ magnetic field,
\begin{align}
\mathcal H_{\Gamma} =& \mathcal H_{AIAO}+\mathcal H_{B,001} \n =& - \alpha (\Gamma_{45} + \tan\phi \cos \theta J_z + \tan\phi \sin \theta J_z^3),
\end{align}
We observe crossing points through perturbation near $\Gamma$ point and projection onto two bands. To complement the argument, we use numerical method to observe crossing points throughout $k$-space. We introduce a pedagogical scheme to observe crossing points.

If we let $\gamma=\tan\phi$ and $\theta = \frac{2\pi}{3}$, then $\gamma$ is the only variable. The energy eigenvalues at $\Gamma$ are
\begin{align}
E_\pm^1 =& \frac{1}{16} ((4-13\sqrt{3})\gamma \pm 2\sqrt{64+(163-53\sqrt{3})\gamma^2}), \n
E_\pm^2 =& \frac{1}{16} ((13\sqrt{3}-4)\gamma \pm 2\sqrt{64+(163-53\sqrt{3})\gamma^2}) \nonumber.
\end{align}
$J_z=\pm3/2, \pm1/2$ are not eigenstates of this Hamiltonian anymore. Since $E_+^1$ and $E_-^2$ are degenerate when $\gamma^* \approx 1.8019$, the energy level sequence changes from $E_-^1<E_+^1<E_-^2<E_+^2$ to $E_-^1<E_-^2<E_+^1<E_+^2$ as $\gamma$ increases. In fact, the sequence exchange between $E_+^1$ and $E_-^2$ at $\gamma^*$ cause the change of the nature of crossing points of two middle bands.

We consider the Luttinger Hamiltonian (Eq. \ref{eq:Luttinger}) as a perturbation to describe the crossing near $\Gamma$. Then, we project the perturbation Hamiltonian onto any a pair of bands. We can establish $6$ possible choices, and observe whether the bands cross or not. Here, we denote $E_1(\vec k)$ to be $E_-^1$, $E_2(\vec k)$ to be $E_-^2$, $E_3(\vec k)$ to be $E_+^1$, $E_4(\vec k)$ to be $E_+^2$, at $\Gamma$ point. Only four choices have crossing points $-$ $E_1$ \& $E_2$, $E_1$ \& $E_3$, $E_2$ \& $E_4$, $E_3$ \& $E_4$, and other choices are gapped.

The projected Hamiltonian has a form like
\begin{align}
H_{proj,\Gamma} = G_0\sigma_0 + G_1\sigma_1 + G_2\sigma_2 + G_3 \sigma_3,
\end{align}
and crossing points are at the solution of $G_1=G_2=G_3=0$. For example, for $E_1$ and $E_2$, one may obtain a system of equations
\begin{align}
(i)\ &k_xk_z = 0 \nonumber \\
(ii)\ &k_yk_z = 0 \nonumber \\
(iii)\ &128\sqrt{3} k_xk_y + \gamma[8(-4+7\sqrt{3})(k_x^2+k_y^2-2k_z^2) \n&+ (4-13\sqrt{3})\eta_1] = 0, \nonumber
\end{align}
where $\eta_1 = \sqrt{64+(163-56\sqrt{3})\gamma^2}$. The solutions are
\begin{align}
(i)\ &k_x = \pm \sqrt{\frac{257}{1048}+\frac{3\sqrt{3}}{131}}\eta_1^{1/2}, k_y = 0, k_z = 0  \nonumber \\
(ii)\ &k_y = \pm \sqrt{\frac{257}{1048}+\frac{3\sqrt{3}}{131}}\eta_1^{1/2}, k_x = 0, k_z = 0. \nonumber \\
(iii)\ &128\sqrt{3} k_xk_y + \gamma[8(-4+7\sqrt{3})(k_x^2+k_y^2)] \n&= \gamma(-4+13\sqrt{3})\eta_1. \nonumber 
\end{align}
The solution $(i)$ and $(ii)$ are, in fact, consistent with the $k_x$ and $k_y$ intersection of $(iii)$, which forms a line node on $k_z=0$ plane. The line node changes its shape as varying $\gamma$ $-$ it is hyperbolic if $\gamma < 1.7055$, a line if $\gamma = 1.7055$, and an ellipse if $\gamma > 1.7055$. 

%%%%%% Energy level crossing %%%%%%%%%%
\begin{figure}
\centering
\includegraphics[width=0.75\columnwidth]{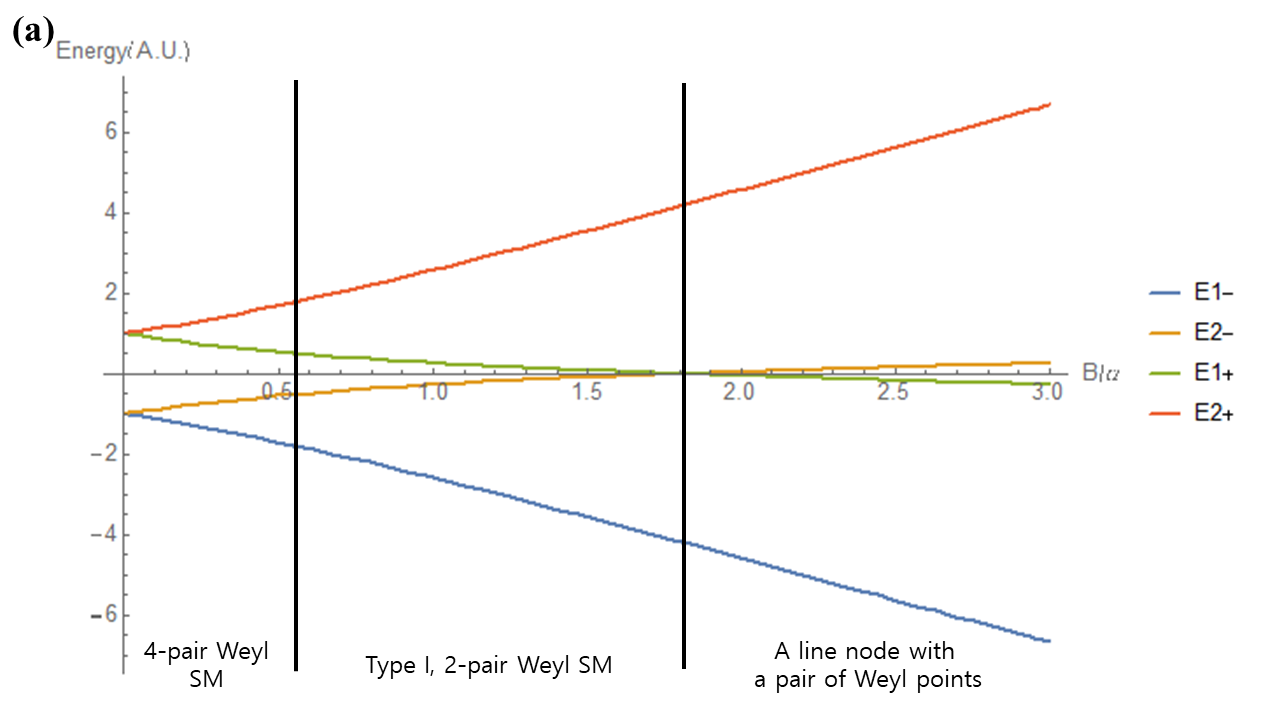}
\includegraphics[width=0.75\columnwidth]{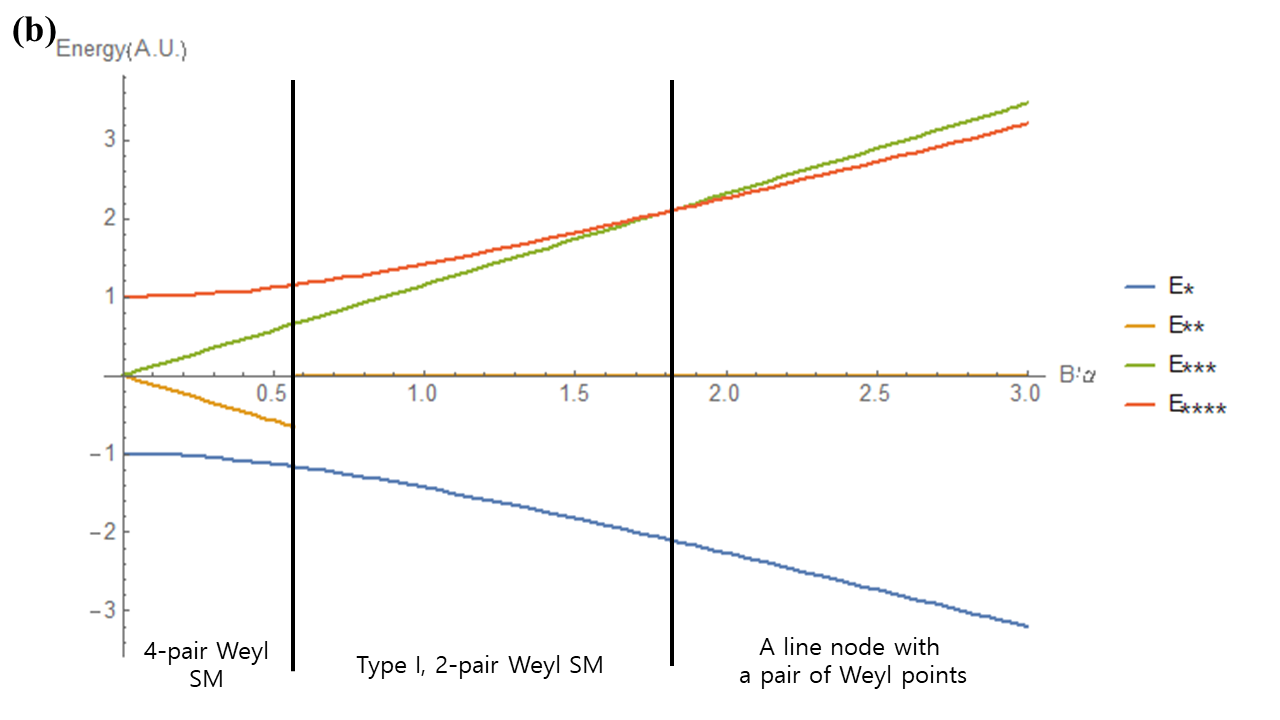}
\includegraphics[width=0.75\columnwidth]{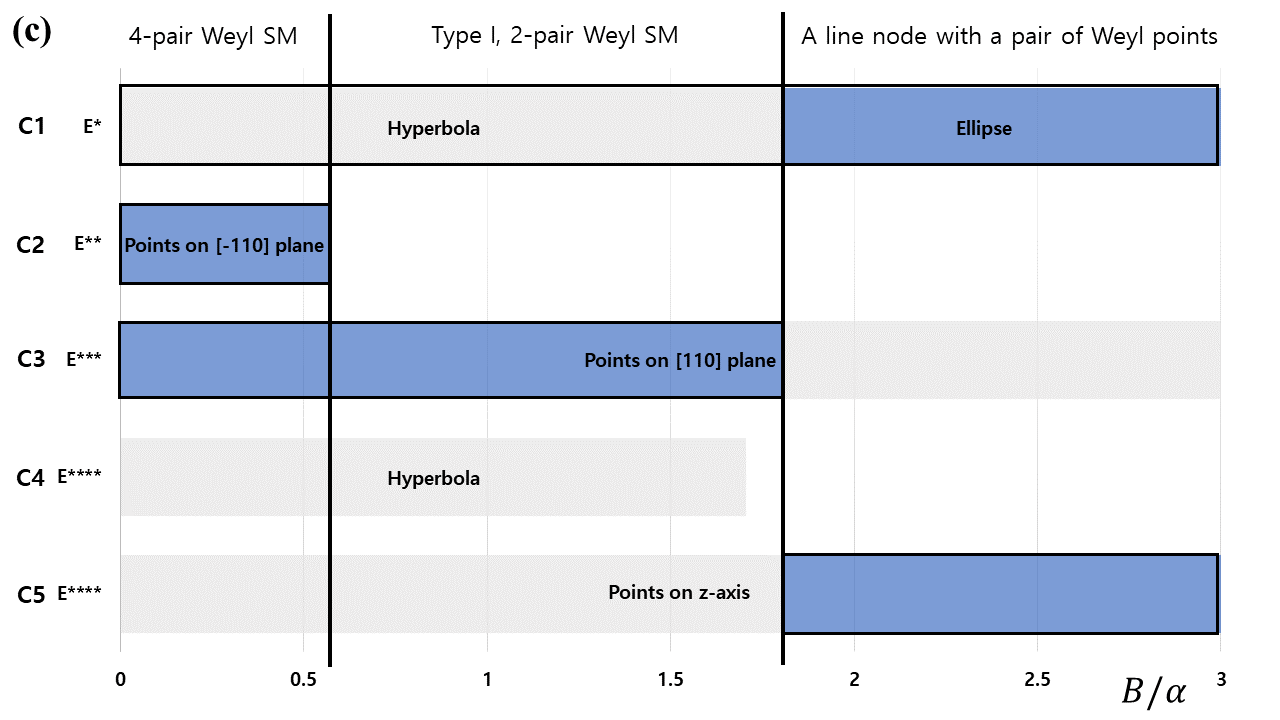}
\caption{(a) Energy spectrums at $\Gamma$ point is drawn by changing $\gamma$. (b) The crossing energy is drawn by changing $\gamma$. We assume the angular momentum unit to be 1, and apply the unit effective Zeeman field. (c) A schematic diagram of 5 classes of crossings is shown against $\gamma$. Each solution class can exist in the blue and gray-shaded regions. The meaning of blue-shaded regions is the crossing between two middle bands. For both figures, the phase diagrams are represented together.}
\label{fig:S2}
\end{figure}

%\begin{figure}
%\centering
%\includegraphics[width=\columnwidth]{Complementary1}
%\caption{The energy dispersion along $k_z$-axis is drawn.}
%\label{fig:Comple1}
%\end{figure}

One can obtain the solutions for other choices with the same way. To sum up, we can classify the solutions into 5 groups.
\begin{enumerate}
\item Crossing between $E_1$ and $E_2$, at energy $E^* = -\frac{1}{8}\eta_1$, whose form is a line node on $k_z=0$ plane; the line node changes from a hyperbola to a line and to an ellipse as $\gamma$ increases. 
\item Crossing between $E_1$ and $E_3$, at energy $E^{**} = -\frac{1}{16}\eta_2$, whose form is 2 pairs of Weyl points on $[1\bar10]$ plane, existing only for $\gamma < 0.5685$. 
\item Crossing between $E_2$ and $E_4$, at energy $E^{***} = \frac{1}{16}\eta_2$, whose form is 2 pairs of Weyl points on $[110]$ plane, existing for every $\gamma$. 
\item Crossing between $E_3$ and $E_4$, at energy $E^{****} = \frac{1}{8}\eta_1$, whose form is a hyperbola on $k_z=0$ plane, only existing for $\gamma < 1.7055$. 
\item Crossing between $E_3$ and $E_4$, at energy $E^{****} =  \frac{1}{8}\eta_1$, whose form is a pair of Weyl points at $k_z$ axis, existing for every $\gamma$.
\end{enumerate}
where $\eta_2 = (13\sqrt{3}-4)\gamma$. 

Although there are various crossings, we concentrate on the crossings between two middle bands repeatedly. Accordingly, we construct a phase diagram by such crossings. In Fig. \ref{fig:S2}, 4-pair Weyl semimetal (4P WSM), Type-1 2-pair Weyl semimetal (T1-2P WSM), and a line-node semimetal (LSM) emerge. T1-2P WSM denotes the phase in which, Group 2 Weyl points in $[1\bar10]$ plane are annihilated while Group 3 Weyl points in $[110]$ plane remain. Remarkably, the phase transition from 4P WSM to 2P WSM is attributed to the annihilation of Weyl points, but the transition from 2P WSM to LSM is come from the energy level sequence exchange between crossing points.

Applying the approach into various $\theta$ and $\gamma = \tan\phi$, we acquire a 2D phase diagram in Fig. \ref{fig:6}. In the phase diagram, in addition to 4P WSM, T1-2P WSM, and LSM, Type-2 2-pair Weyl semimetal(T2-2P WSM) and Double Weyl semimetal(DWSM) emerge. T2-2P WSM is the phase in which Group 2 points remain while Group 3 points vanish. The phase transition from 2P WSM to DWSM emerges from merging a pair of Weyl points with the same topological charge at $k_z$-axis.

In summary, the result implies that diverse topological phases can arise by changing $\theta$ and $\phi$, and which phase transition occurs depends heavily on the interplay between Zeeman and Luttinger q-term. We turn out that for a certain range of $\theta$ ($\theta_{b1} < \theta <\theta_{b2}, \theta_{b3} < \theta < \theta_{b4}$), LSM appears, while DWSM emerges for the remaining range. For DWSM and LSM, not only are the shape and positions of crossings different, but also the way of phase transition is disparate. The phase transition from 2P WSM to DWSM occurs by the combination of Weyl points at high-symmetry line. while the transition from 2P WSM to LSM is from the exchange of energy level sequence at $\Gamma$ point changes two middle bands.

%%%%%%%%%%%%%% Trajectory 111 %%%%%%%%%%%%%%%%
\begin{figure}[b]
\includegraphics[width=\columnwidth]{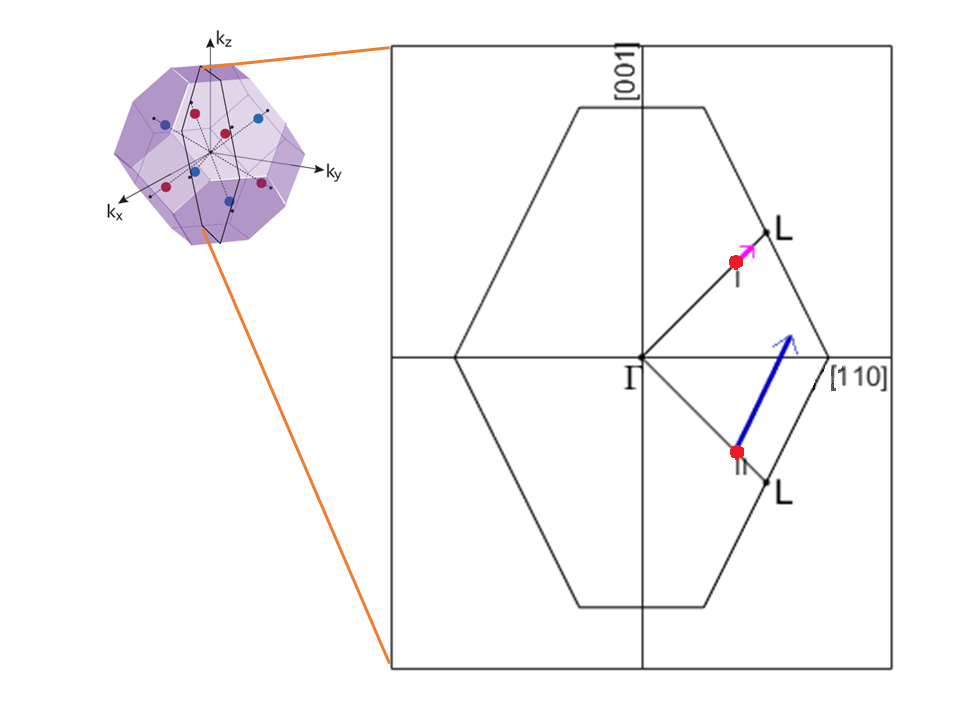}
\caption{Under $[111]$ effective field, the trajectories of Weyl points are drawn in colored lines when Zeeman and Luttinger q-term are equally contributed ($\beta_1=\beta_2$). Class 1 Weyl point in a pink line is never deviated from $[111]$ axis, while Class 2 Weyl point in a blue line moves away from the symmetry line, but still remains on the mirror plane. There is no zone boundary in this model, yet it is drawn in order to visualize the Weyl points effectively.}
\label{fig:Trajectory}
\end{figure}

%111 direction
\subsubsection{[111] direction \label{[111]}}

Under $[111]$ direction field, we begin to study the trajectories of 8 Weyl points under large AIAO order parameter and small magnetic field. Since 3-fold rotation around $[111]$ line $C_{3,111}$ is still preserved, 8 Weyl points will be divided into 2 classes again. Class 1 includes 2 Weyl points along [111] line, while Class 2 does other 6 Weyl points. Class 1 points will never be deviated from $[111]$ line, and Class 2 points will travel only on the mirror planes, according to the symmetries and the topological nature of Weyl points.

For Class 1, let us choose $\vec k_{C1,111}=\sqrt{\frac{\alpha}{3}} (1,1,1)$. The Hamiltonian near the Weyl points is just
\begin{align}
\mathcal H_{C1}^{111} = \mathcal H_{C1}^{001},
\end{align}
while the magnetic field Hamiltonian is Eq. \ref{eq:111}.

After the same procedure as the analysis of $[001]$ direction, we have
\begin{align}
\mathcal H_{proj,C1}^{111} =& A_0 \sigma_0 + A_1 \sigma_x  + A_2 \sigma_y + A_3\sigma_z,
\end{align}
where
\begin{align}
A_0 =& \frac{\sqrt{3}}{2}\beta_1 + \frac{3}{2}\frac{\beta_1^2}{\alpha} + \frac{13\sqrt{3}}{8}m_2 + \frac{147}{32}\frac{m_2^2}{\alpha}\nonumber \\
A_1 =& -\frac{3}{2\sqrt{2}}\frac{\beta_1^2}{\alpha}-\frac{147}{32\sqrt{2}}\frac{\beta_2^2}{\alpha} \n&-\sqrt{\alpha}[\frac{-(2-\sqrt{3})q_x-(2+\sqrt{3})q_y-2q_z}{\sqrt{6}}]\nonumber \\
A_2 =& \frac{3}{2\sqrt{2}}\frac{\beta_1^2}{\alpha}+\frac{147}{32\sqrt{2}}\frac{\beta_2^2}{\alpha} \n&-\sqrt{\alpha}[\frac{(2+\sqrt{3})q_x+(2-\sqrt{3})q_y+2q_z}{\sqrt{6}}]\nonumber \\
A_3 =& \frac{\sqrt{\alpha}(q_x+q_y-2q_z)}{\sqrt{3}} .
\end{align}
Weyl points exist at the solutions of $A_1=A_2=A_3=0$.
\begin{align}
(q_x,q_y,q_z) = \frac{\sqrt{3}}{64}\frac{16\beta_1^2+49\beta_2^2}{\alpha^{3/2}}(1,1,1).
\end{align}
According to the inversion symmetry, another Weyl point in Class 1 moves to $(q_x,q_y,q_z) = -\frac{\sqrt{3}}{64}\frac{16\beta_1^2+49\beta_2^2}{\alpha^{3/2}}(1,1,1)$. Class 1 Weyl points stick on $[111]$ line.

On the other hand, if we choose one of Class 2 Weyl point, $\vec k_{C2,111} = \sqrt{\f{\alpha}{3}}(1,1,-1)$, the Hamiltonian is 
\begin{align}
\mathcal H_{C2}^{111}=\mathcal H_{0,C2}^{111}+\mathcal H_{mom,C2}^{111},
\end{align}
where
\begin{align}
\mathcal H_{0,C2}^{111} =& \frac{\alpha}{\sqrt{3}}(\Gamma_1 + \Gamma_2 - \Gamma_3) - \alpha\Gamma_{45} \nonumber \\
\mathcal H_{mom,C2}^{111}=&  -\sqrt{\alpha}[(q_z-q_y)\Gamma_1 + (q_z-q_x)\Gamma_2 \n&+ (q_x+q_y)\Gamma_3 + (q_x-q_y)\Gamma_4 \n&+ \frac{1}{\sqrt{3}}(-2q_z-q_x-q_y)\Gamma_5],
\end{align}
up to first order of $\vec q = \vec k - \vec k_{C2,111}$. The magnetic field Hamiltonian is Eq. \ref{eq:111}, again.

Given by the same procedure, the Class 2 Weyl point will be at
\begin{align}
q_x =&~ q_y = \frac{-16\alpha(4\beta_1+7\beta_2)+\sqrt{3}(48\beta_1^2+151\beta_2^2)}{192\alpha^{3/2}} \nonumber \\
q_z =&~ \frac{-32\alpha(4\beta_1+7\beta_2)+\sqrt{3}(48\beta_1^2+395\beta_2^2)}{192\alpha^{3/2}}.
\end{align}
For other 5 Weyl points, $C_{3,111}$ and $P$ determine the trajectories. Class 2 Weyl points are deviated away from the high-symmetry axes due to the 3-fold rotational symmetry breaking, but the points cannot travel out of the mirror planes by $\sigma_{d,111}T$ symmetries. If Class 2 Weyl points move out of the plane, Nielsen-Ninomiya Theorem is violated. The trajectories of both classes of Weyl points are drawn in Fig. \ref{fig:Trajectory}.

\begin{figure}[b]
\centering
\includegraphics[width=\columnwidth]{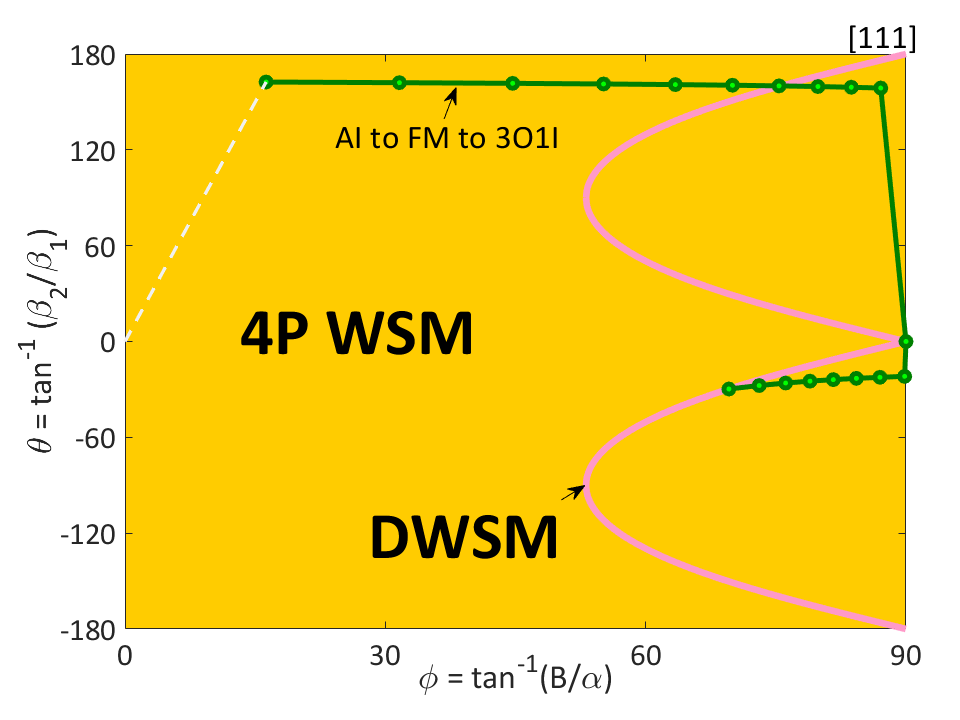}
\caption{A general 2D phase diagram under $[111]$ direction effective field is drawn. A couple of topological phases, 4-pair Weyl semimetal (4P WSM), Double Weyl semimetal (DWSM) can be observed. Green line shows the projection of Hubbard interaction, from AI to FM to 3O1I.}
\label{fig:2DPD111}
\end{figure}

Next, we investigate the crossing point by varying $\theta$ and $\phi$. The Hamiltonian is
\begin{align}
\mathcal H =\mathcal H_0 + \mathcal H_{AIAO}+\mathcal H_{B,111}.
\end{align}
According to the previous section, double Weyl points emerge only if $J_{[111]}=\pm 3/2$ are eigenstates of the Hamiltonian. Finding the condition that $J_{[111]}$ eigenstates diagonalize the Hamiltonian $H$, one can obtain a line for DWSM phase. In Fig. \ref{fig:2DPD111}, we represent a general phase diagram under $[111]$ direction of effective field.

%%%%%%%%%%%%%% 2D phase diagram 111 %%%%%%%%%%%%%

In a nutshell, we observe a number of distinct topological phases under effective field: DWSM, 4P WSM, T1/T2-2P WSM, and LSM. The interplay between diverse magnetic terms play an important role on the emergence of distinct topological phases.

%******************************* Effective Theory at L ******************************%
\section{Effective Theory at $L$ point \label{sec:App2}}

According to previous research\cite{witczak2012topological}, if AIAO order parameter is developed, pyrochlore iridates become the insulating phase. In order to observe topological phases near the insulating phase, we should study the effective theory near $L$ point. $L$ points have lower symmetry than $\Gamma$ point $-$ only $T$, $P$, $C_3$, and some of $\sigma_d$ are preserved.

%********** General Hamiltonian at L point
\subsection{General Hamiltonian at $L$ point}

By the inversion symmetry $P$ at $L$ point, energy eigenstates at $L$ point must have either one of $P$ eigenvalues, $\pm1$ ($P^2=1$). We choose two eigenstates with distinct eigenvalues. If we take $P=\sigma_z$, the 2-band Hamiltonian at $L$ point will be
\begin{align}
\mathcal H_L = \Delta \sigma_z.
\end{align}
Let us define the local $z$-direction be along the 3-fold rotation axis, and local $y$-direction be in the mirror plane. Near $L$ point, the most general Hamiltonian which is invariant under $P$ symmetry up to second order of momentum $\vec q=\vec k-\vec k_L$ is
\begin{align}
\mathcal H_1(\vec{q}) =& A_1(\vec q)\sigma_x + B_1(\vec q)\sigma_y + (C_1(\vec q)+D_1(\vec q))\sigma_z.
\end{align}
such that
\begin{align}
A_1(\vec q) =&~ a_1q_x+a_2q_y+a_3q_z \nonumber \\
B_1(\vec q) =&~ b_1q_x+b_2q_y+b_3q_z \nonumber \\
C_1(\vec q) =&~ \Delta + \frac{q_x^2}{2m_x} + \frac{q_y^2}{2m_y} + \frac{q_z^2}{2m_z} \nonumber \\
D_1(\vec q) =&~ c_1q_xq_y + c_2q_xq_z+c_3 q_yq_z. \nonumber
\end{align}
Next, we impose $\sigma_d T$ symmetry upon this Hamiltonian. Considering that $\sigma_dT$ is anti-unitary, $(\sigma_dT)^2=1$, and $\Delta \sigma_z$ is invariant under the symmetry, one can choose $\sigma_dT=K$ (complex conjugate). The general Hamiltonian near $L$ point under $P$ and $\sigma_d T$ is
\begin{align}
\mathcal H_2(\vec{q}) =& A_2(\vec{q})\sigma_x + B_2(\vec q)\sigma_y + (C_2(\vec q) + D_2 (\vec q))\sigma_z. \label{eq:H2}
\end{align}
where
\begin{align}
A_2(\vec q) =&~ a_1q_x \nonumber \\
B_2(\vec q) =&~ b_2q_y+b_3q_z \nonumber \\
C_2(\vec q) =&~ \Delta + \frac{q_x^2}{2m_x} + \frac{q_y^2}{2m_y} + \frac{q_z^2}{2m_z} \nonumber \\
D_2(\vec q) =&~ c_3q_yq_z \nonumber
\end{align}
Finally, we add up 3-fold rotation symmetry about local $z$-axis, $C_3 = e^{i\frac{2\pi}{3}\sigma_z}$. The general Hamiltonian under $L$ point under $P$, $\sigma_dT$, and $C_3$;
\begin{align}
\mathcal H_3(\vec{q}) =& a(q_x\sigma_x+ q_y\sigma_y) \n& + (\Delta + \frac{q_x^2+q_y^2}{2m_{xy}} + \frac{q_z^2}{2m_z})\sigma_z \label{eq:H3}
\end{align}
$\mathcal H_2$ is the most general Hamiltonian with $P$ and $\sigma_dT$, while $\mathcal H_3$ is the most general Hamiltonian with $P$, $\sigma_dT$, and $C_{3}$.

We can establish the general Hamiltonian with effective field up to first order under $P$ and $\sigma_d T$, as well.
\begin{align}
\mathcal H_{2B}(\vec q, \vec B) = A_{2B}\sigma_x + B_{2B} \sigma_y + C_{2B} \sigma_z,
\end{align}
where
\begin{align}
A_{2B} =& q_x(d_1B_y+d_2B_z)+B_x(d_3q_y+d_4q_z) \nonumber \\
B_{2B} =& B_y(e_1q_y+e_2q_z)+B_z(e_3q_y+e_4q_z) \nonumber \\
C_{2B} =& f_1B_y+f_2B_z. \nonumber
\end{align}
Adding $C_3$ symmetry, one can find out the Hamiltonian with magnetic field.
\begin{align}
\mathcal H_{3B}(\vec q, \vec B) = gB_z(q_x\sigma_x + q_y\sigma_y) + f_2B_z\sigma_z.
\end{align}

% 111 direction
\subsection{$[111]$ direction}

Here, we begin with the phase of [111] direction of effective field since we can understand the phase of $[001]$ direction through the argument in this section. 

We divide all of 4 $L$ points in Brillouin zone into 2 classes $-$ Class 1 $L$ point is an $L$ point on [111]-axis, while Class 2 $L$ points are other three. Without magnetic field, the Hamiltonian is just Eq. \ref{eq:H3} for every $L$ point. We obtain the position of Weyl point as
\begin{align}
q_x=q_y=0,\ q_z=\pm\sqrt{-2m_z\Delta} \nonumber 
\end{align}
A pair of Weyl points exist along local $z$-axis only if $m_z\Delta < 0$. 

If we apply the magnetic field on the system, every symmetry of Class 1 $L$ point remains preserved.
\begin{align}
\mathcal H_{111}^{L,C1} =& \mathcal H_3 +\mathcal H_{3B} = a'(q_x\sigma_x+q_y\sigma_y) \n&+ (\Delta' + \frac{q_x^2+q_y^2}{2m_{xy}} + \frac{q_z^2}{2m_z})\sigma_z \label{eq:HC1111}
\end{align}
where
\begin{align}
a' =& a+gB_z ,~\Delta' = \Delta + f_2B_z. \nonumber
\end{align}
Since the form of Eq. \ref{eq:HC1111} is the same as Eq. \ref{eq:H3}, the positions of Weyl points are just $(q_x,q_y,q_z) = (0,0,\pm \sqrt{-2m_z\Delta'})$. That is, Weyl points can only move along local $z$-axis, which corresponds to global [111] line. Furthermore, if $-m_z\Delta'=0$, two Weyl points meet at the origin, and if $-m_z\Delta'>0$, the pair of Weyl points are annihilated. If $m_z>0$, then the condition for the gapless state is $\Delta' = \Delta + f_2B_z<0$.

%%%%%%%%%%%%% L Phase 111 %%%%%%%%%%%%%
\begin{figure}[t]
\includegraphics[width=\columnwidth]{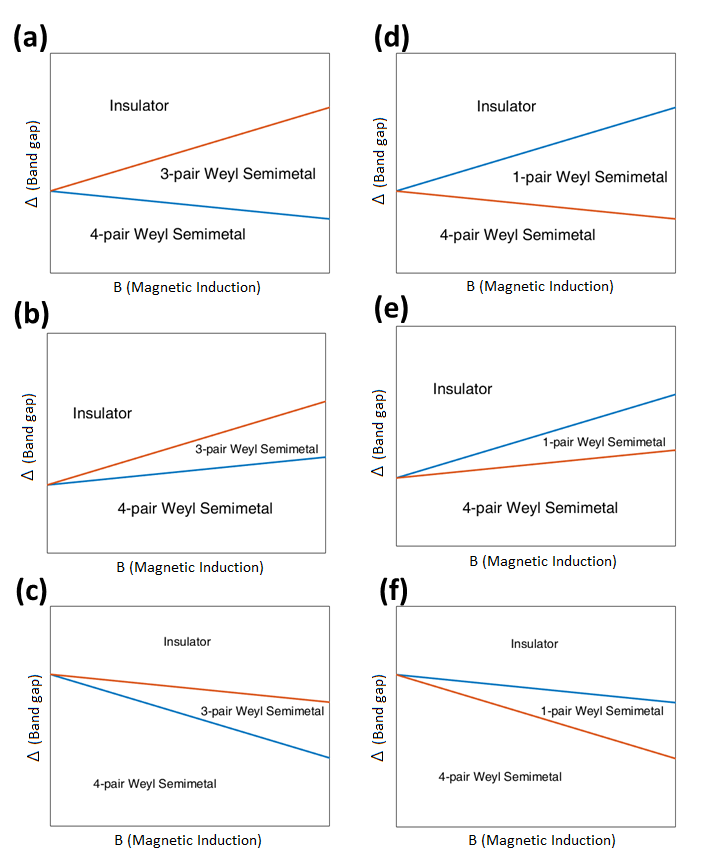}
\caption{The possible phase diagrams near insulating phase of pyrochlore iridates under [111] direction of effective field are shown. The blue lines are where the Class 1 $L$ points become gapless, while the orange lines are where the Class 2 $L$ points become gapless. The slope of each line is (a) $f_2 = 0.1, f_2'= -0.3$, (b) $f_2 = -0.3, f_2' = 0.1$, (c) $f_2 = -0.1, f_2'= -0.3$, (d) $f_2 = -0.3, f_2'=-0.1$, (e) $f_2 = 0.3, f_2'=0.1$, and (f) $f_2 =0.1 , f_2'=0.3$.}
\label{fig:LPhase111}
\end{figure}

%%%%%%%%%%%%%%%%%%%%%% L Phase 001 %%%%%%%%%%%%%%%%%%%%%%%
\begin{figure}[b]
\centering
\includegraphics[width=\columnwidth]{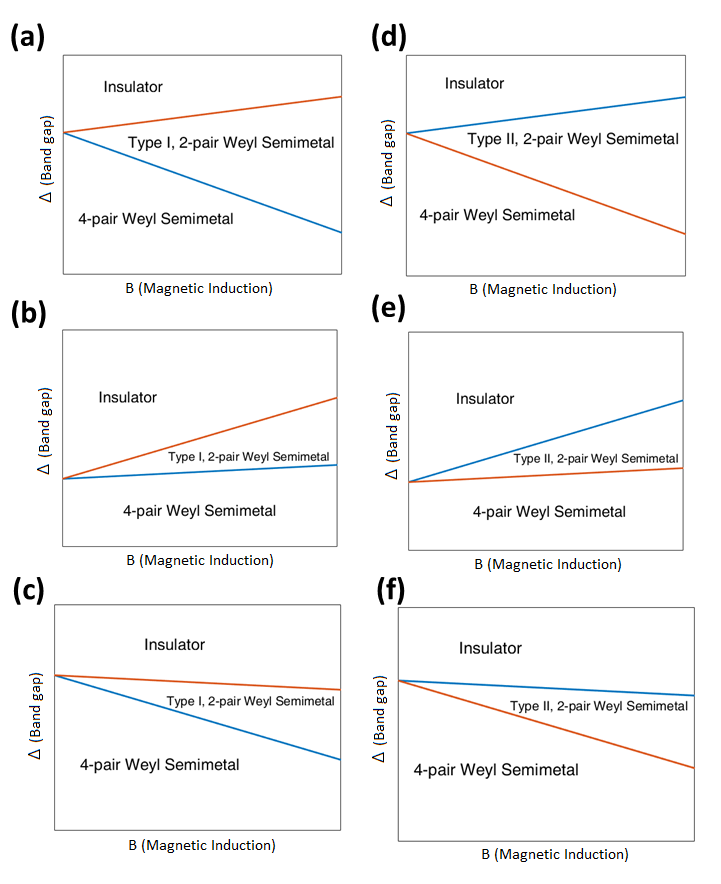}
\caption{The possible phase diagrams near Weyl semimetal-Insulating phase transition under [001] direction field are shown. The blue lines are where the Class 1 $L$ points become gapless, while the orange lines are where the Class 2 $L$ points become gapless. The slope of each line is (a) $f_1'= 0.2, f_2' =0.2$,(b) $f_1'= 0.1, f_2' =-0.3$, (c) $f_1'= 0.1, f_2' =0.3$, (d) $f_1'= -0.2, f_2' =0.2$, (e) $f_1'= -0.1, f_2' =-0.3$, (f) $f_1'= -0.1, f_2' =0.3$.}
\label{fig:LPhase001}
\end{figure}

At Class 2 L points, $C_3$ symmetry is broken. 
\begin{align}
\mathcal H_{111}^{L,C2}=& \mathcal H_2 + \mathcal H_{2B} \n=&  A_2'(\vec{q})\sigma_x + B_2'(\vec q)\sigma_y +(C_2'(\vec q)+D_2'(\vec q)) \sigma_z \label{eq:C2111}
\end{align}
where
\begin{align}
A_2(\vec q) =& q_x(a_1+d_1B_y+d_2B_z) = a_1'q_x\nonumber \\
B_2(\vec q) =& q_y(b_2+e_1B_y+e_3B_z)+q_z(b_3 + e_2B_y+e_4B_z) \n=& b_2'q_y+b_3'q_z \n
C_2(\vec q) = &(\Delta + f_1B_y+f_2B_z)+ \frac{q_x^2}{2m_x} + \frac{q_y^2}{2m_y} + \frac{q_z^2}{2m_z} \n=& \Delta'' + \frac{q_x^2}{2m_x} + \frac{q_y^2}{2m_y} + \frac{q_z^2}{2m_z} \nonumber \\
D_2(\vec q) =& c_3q_yq_z
\end{align}
Note that $B_x = 0$ here, since the magnetic field direction is in $k_x=0$ plane for Class 2 $L$ points. Without magnetic field, we should obtain $H_{3B}$ again, so that $b_2=a_1$, $b_3 = c_3 =0$, and $m_x = m_y = m_{xy}$. The Hamiltonian Eq. \ref{eq:C2111} is just the renormalization of some variables in Eq. \ref{eq:H2}. Weyl points will exist at
\begin{align}
q_x =& 0,\ q_y = \pm \frac{b_3'}{b_2'}\sqrt{-\Delta''(\frac{b_3'^2}{2m_{xy}b_2'^2}+\frac{1}{2m_z})},\n q_z =& \pm \sqrt{-\Delta''(\frac{b_3'^2}{2m_{xy}b_2'^2}+\frac{1}{2m_z})} \nonumber
\end{align}
A pair of Weyl points exist only if $-\Delta''(\frac{b_3'^2}{2m_{xy}b_2'^2}+\frac{1}{2m_z})\equiv -\Delta'' X>0$, and the pair annihilation occurs at origin if $\Delta'' = 0$. If we assume $X>0$, Weyl points exist when $\Delta'' < 0$. Near Class 2 $L$ points, Weyl points can move off from the high symmetry line and travel through local $yz$ mirror plane. The result is consistent with the trajectory in $\Gamma$ effective theory of Sec. \ref{[111]}.

In summary, we have two equations to obtain phase transitions. 
\begin{align}
\Delta + f_2 B = 0,\ \Delta + f_2'B = 0, \nonumber
\end{align}
Usually, $f_2$ and $f_2'$ does not have to be equal to each other. In Fig. \ref{fig:LPhase111}, we represent all possible forms of phase diagram by changing $f_2$ and $f_2'$. One can observe 4-pair Weyl semimetal (4P WSM), 3-pair Weyl semimetal (3P WSM), 1-pair Weyl semimetal (1P WSM) and trivial insulator (INS). Which topological semimetal emerge depends on the sequence of Weyl point annihilation. If Weyl points are annihilated at Class 1 $L$ point first, we can observe 3-pair Weyl semimetal, while if Weyl points are annihilated at Class 2 $L$ points first, we can observe 1-pair Weyl semimetal.

%% 001 direction
\subsection{$[001]$ direction}

Under $[001]$ direction of effective field, we divide 4 $L$ point into into 2 classes again $-$ a pair of $L$ points in $[1\bar10]$ plane are Class 1, and another pair of $L$ points in $[110]$ plane are Class 2. Since 3-fold rotational symmetries are broken while $2 \sigma_dT$ remains, both classes are just the same as Class 2 Weyl points of $[111]$ case. Again, we set local $z$-axis along 3-fold rotation axis, and local $y$-axis inside the mirror plane. Recalling Eq. \ref{eq:C2111} and the solutions, we confront two following equations as well. 
\begin{align}
\Delta+ f_1'B = 0,\ \Delta+ f_2'B = 0 \nonumber
\end{align}
By controlling $f_1'$ and $f_2'$, we draw several forms of phase diagrams in Fig. \ref{fig:LPhase001}.

In the phase diagram, we observe Type I and II 2-pair Weyl semimetal. T1-2P WSM denotes the semimetal without Weyl points near Class 1 $L$ points, while T2-2P WSM denotes that without Weyl points near Class 2 $L$ points. The sequence of Weyl point annihilation determines topological semimetallic phase. If Class 1/2 Weyl points are annihilated initially, then T1/2-2P WSM appears.

In summary, we can observe the emergent topological phases like 3P WSM, T1/2-2P WSM, and 1P WSM near insulating phase.

\section{Cluster Magnetic Multipole in Pyrochlore Iridates \label{sec:App3}}

\begin{figure}[b]
\centering
\includegraphics[width=\columnwidth]{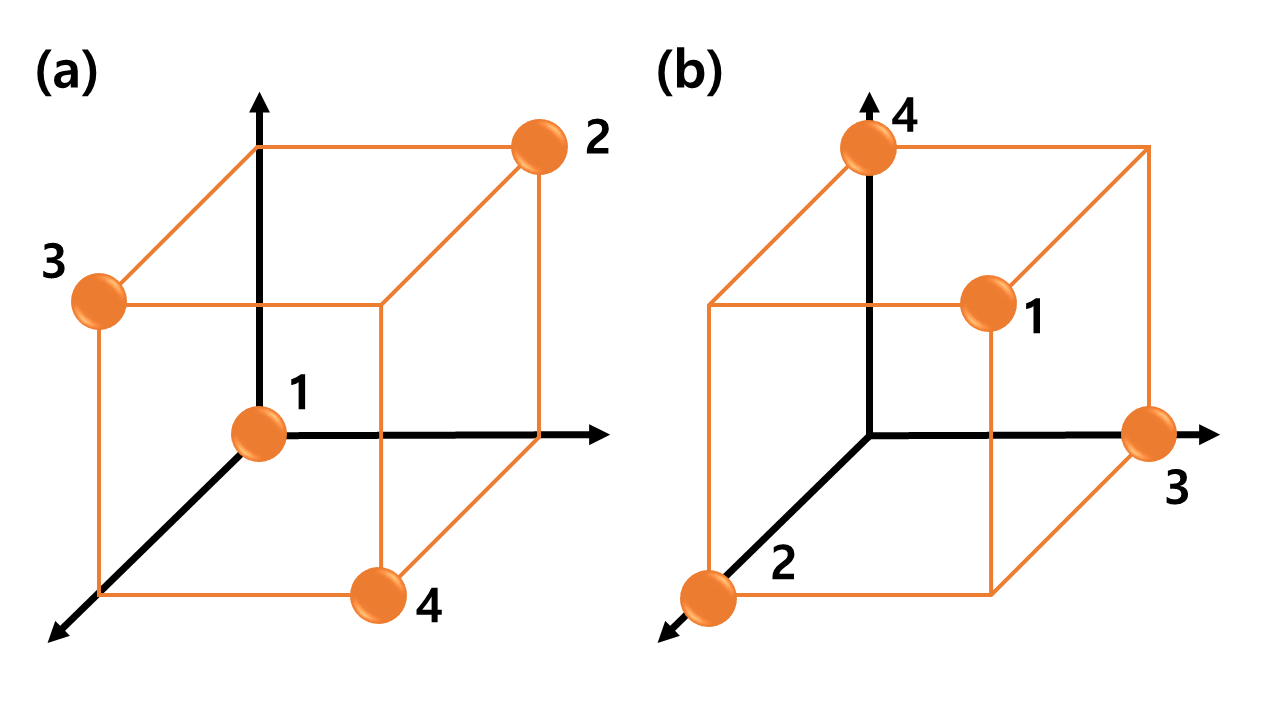}
\caption{Two clusters in pyrochlore iridates are shown. They are related by nonsymmorphic symmetry $\{P|T_{1/4,1/4,1/4}\}$.}
\label{fig:cluster}
\end{figure}

\subsection{Cluster Magnetic Multipoles}

Suppose we have a piece of magnetic matter localized in the real space. Then, Ampere-Maxwell's law becomes
\begin{align}
\grad^2 \V A = \mu_0\V J
\end{align}
outside the matter, under Coulomb gauge ($\div \V A = 0$). By Green's theorem, we obtain the general solution as
\begin{align}
\V A = \sum_{p=0}^{\infty} \sum_{q=-p}^{p} \f{1}{r^{p+1}} \V Z_{pq}(\theta,\phi) M_{pq},
\end{align}
where 
\begin{align}
\V Z_{pq} =& -\f{i}{p}\V L [\sqrt\f{4\pi}{2p+1}Y_{pq}(\theta,\phi) ], \n
M_{pq} =&\sqrt\f{4\pi}{2p+1} \int d^3 r'\ \grad'[r'^pY^*_{pq}(\theta',\phi') ]\cdot \V M(\V r'), \label{eq:mag}
\end{align}
such that $\V L$ is angular momentum, $Y_{pq}(\theta,\phi)$ is spherical harmonics, and $\V M(\V r')$ is the magnetization density defined by $\V J(\V r') = c\curl \V M(\V r')$. This process is called multipole expansion, and $M_{pq}$ is called magnetic multipole. In general, we can express any configurations of magnetic matter into the series of multipoles \cite{kusunose2008description}.

%%%%%%%%%%%%%%%%%Table CMMM multipole%%%%%%%%%%%%%%%

Applying the same argument in the lattice, we can define cluster magnetic multipole moment (CMMM) \cite{suzuki2017cluster}. An atom cluster is defined as a group of atoms connected by point group operators within a magnetic unit cell. CMMM at $a$-th cluster in the magnetic unit cell is simply defined as same as Eq. \ref{eq:mag},
\begin{align}
M_{pq}^{a} =&\sqrt\f{4\pi}{2p+1} \sum_{i=1}^{N_a} \grad[r_i^p Y^*_{pq}(\theta_i,\phi_i) ] \cdot \V m_i, \label{eq:CMMM}
\end{align}
where $\V m_i$ is the magnetic moment at $i$-th site, and $N_a$ is the number of atoms in $a$-th cluster. This is a spherical tensor of rank $p$. If $p=1,2,$ and $3$, then we can acquire the dipoles, quadrupoles, and octupoles of the cluster, respectively. The contribution of the magnetic unit cell on CMMM is just the summation over every cluster in the cell,
\begin{align}
M_{pq} = \f{N^u}{N^c}\f{1}{V}\sum_{a=1}^{N} M_{pq}^a.
\end{align}
where $N^u$ is the number of atoms of the magnetic unit cell, $N^c$ is the total number of atoms in every cluster, $N$ is the number of clusters, $V$ is the volume of the magnetic unit cell.

%%%%%%%%%%%%%%%%%CMMM multipole%%%%%%%%%%%%%%%%
\begin{figure*}[ht]
\centering
\includegraphics[width=2\columnwidth]{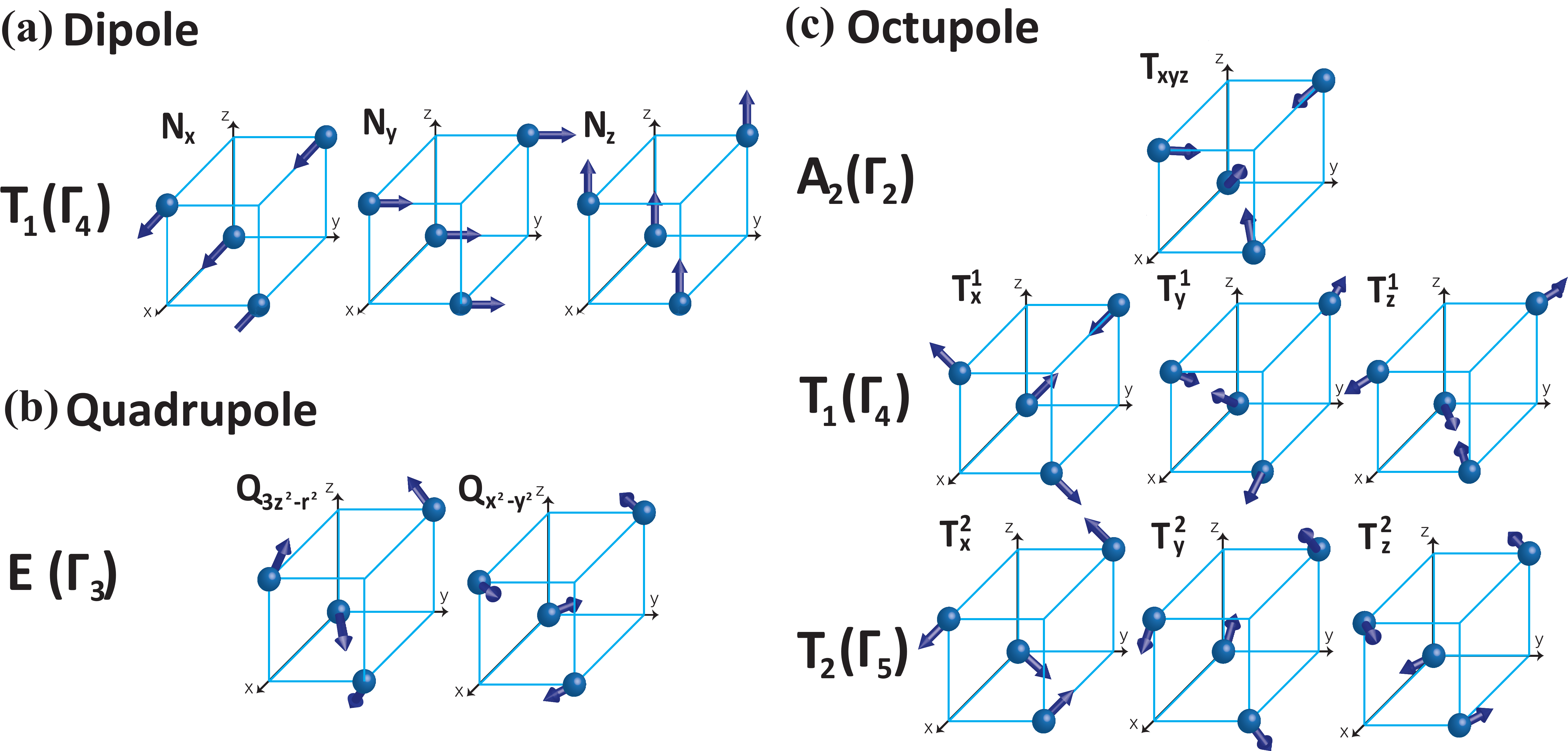}
\caption{Each symmetrized CMMM in TABLE \ref{tab:CMMM} is depicted in the first cluster of pyrochlore iridates with its group representation. The configuration of $T_x^1, T_y^1, T_z^1$ are just the same as that of $Q_{yz}, Q_{zx}, Q_{xy}$. Due to the inversion symmetry, quadrupole will be cancelled by the second cluster.}
\label{fig:CMMM}
\end{figure*}

\subsection{CMMM in Pyrochlore Iridates}

If we consider the magnetic order of the wavevector $\V q=0$, an magnetic unit cell is just the same as a unit cell and an atomic cluster. We assume the length of unit cell edge is 1.

There are two clusters, which are related by nonsymmorphic symmetry operation $\{P|T_{1/4,1/4,1/4}\}$ (See. Fig. \ref{fig:cluster}). In a cluster, the number of degree of freedom is twelve, since there are 3 moment directions and 4 atomic sites. Accordingly, we expect that dipoles, quadrupoles, and octupoles appear in the cluster, since the number of CMMM components is fifteen up to octupoles. We denote $\alpha=x,y,z$ component of the magnetic moment of $i$-th site as $m_{i\alpha}$. Then, using Eq. \ref{eq:CMMM}, the cluster dipoles are
\begin{widetext}
\begin{align}
M_{1\bar1} =& \f{1}{\sqrt{2}}[m_{1x}+ i m_{1y} + m_{2x}+ i m_{2y} + m_{3x}+ i m_{3y} + m_{4x}+ i m_{4y}], \n
M_{10} =& m_{1z}+m_{2z}+m_{3z}+m_{4z}, \n
M_{11} =& \f{1}{\sqrt{2}}[- m_{1x}+ i m_{1y} - m_{2x}+ i m_{2y} - m_{3x}+ i m_{3y} - m_{4x}+ i m_{4y}]. \label{eq:dipo}
\end{align}
\end{widetext}
The cluster quadrupoles for a cluster exist, but the total cluster quadrupoles are all zero, since the system is inversion-symmetric while quadrarupoles aren't ($M_{pq}=(-1)^{p+1}M_{pq}$). Each cluster can have quadrupole moments; for example, Cluster 1 (Fig. \ref{fig:cluster}a) has
\begin{widetext}
\begin{align}
M_{2\bar2}^{(1)} =& \f{\sqrt{3}}{2}[ e^{-i\f{3\pi}{4}}(m_{1x} + im_{1y}) + e^{i\f{3\pi}{4}}(m_{2x} + im_{2y}) + e^{-i\f{\pi}{4}} (m_{3x} + im_{3y}) +e^{i\f{\pi}{4}}(m_{4x} + im_{4y})], \n
M_{2\bar1}^{(1)} =&  \f{\sqrt{3}}{2}[ \f{1}{\sqrt{2}}(- m_{1x} - im_{1y}+ m_{2x} + im_{2y} + m_{3x} + im_{3y} - m_{4x} - im_{4y} )+ e^{-i\f{3\pi}{4}}m_{1z} + e^{i\f{3\pi}{4}}m_{2z}  + e^{-i\f{\pi}{4}}m_{3z} + e^{i\f{\pi}{4}} m_{4z}],\n
 M_{20}^{(1)} =& \f{1}{2} [m_{1x}+m_{1y}-2m_{1z}+m_{2x}-m_{2y}+2m_{2z}-m_{3x}+m_{3y}+2m_{3z} - m_{4x}-m_{4y}-2m_{4z}], \n
   M_{21}^{(1)} = & \f{\sqrt{3}}{2}[ \f{1}{\sqrt{2}}(m_{1x} - im_{1y} - m_{2x} + im_{2y} - m_{3x} + im_{3y} + m_{4x} - im_{4y} )+ e^{-i\f{\pi}{4}}m_{1z} + e^{i\f{\pi}{4}}m_{2z}  + e^{-i\f{3\pi}{4}}m_{3z}+ e^{i\f{3\pi}{4}} m_{4z}],\nonumber
 \end{align}
 \begin{align}
  M_{22}^{(1)} =& \f{\sqrt{3}}{2}[ e^{i\f{3\pi}{4}}(m_{1x} - im_{1y}) + e^{-i\f{3\pi}{4}}(m_{2x} - im_{2y}) + e^{i\f{\pi}{4}} (m_{3x} - im_{3y}) + e^{-i\f{\pi}{4}}(m_{4x} - im_{4y})]. \label{eq:quad}
 \end{align}
 \end{widetext}
and these are cancelled out by the quadrupole moments of Cluster 2. By the way, the cluster octupoles are
\begin{widetext}
\begin{align}
M_{3\bar3} =& \f{3\sqrt{5}}{8} i[ m_{1x} + im_{1y} - m_{2x} - im_{2y} - m_{3x} - im_{3y} + m_{4x} + im_{4y}] \n
M_{3\bar2} =&\f{\sqrt{15}}{4}[ e^{i\f{\pi}{4}}(m_{1x}-m_{2y}+m_{3y}-m_{4x}) + e^{i\f{3\pi}{4}} (m_{1y}+m_{2x}-m_{3x}-m_{4y}) +\f{i}{\sqrt{2}}(m_{1z} -m_{2z} - m_{3z} + m_{4z})] ,\n
M_{3\bar1} =& \f{\sqrt{3}}{8}[-i(m_{1x}-m_{2x}-m_{3x}+m_{4x}) -( m_{1y} -m_{2y} - m_{3y}+ m_{4y}) +4\sqrt{2}\{e^{i\f{\pi}{4}}(m_{1z}-m_{4z}) + e^{i\f{3\pi}{4}}(m_{2z} - m_{3z})\}] ,\nonumber \\
M_{30} =& \f{3}{4}[-m_{1x}-m_{1y}+m_{2x}-m_{2y}-m_{3x}+m_{3y}+m_{4x}+m_{4y}], \n
M_{31}=& \f{\sqrt{3}}{8}[-i(m_{1x}-m_{2x}-m_{3x}+m_{4x})+(m_{1y}-m_{2y}-m_{3y}+m_{4y})+4\sqrt{2}\{e^{i\f{3\pi}{4}}(m_{1z}-m_{4z}) + e^{i\f{\pi}{4}}(m_{2z} - m_{3z})\}], \nonumber \\
M_{32} =& \f{\sqrt{15}}{4}[ e^{-i\f{\pi}{4}}(m_{1x}-m_{2y}+m_{3y}-m_{4x}) + e^{-i\f{3\pi}{4}}(m_{1y}+m_{2x}-m_{3x}-m_{4y}) -\f{i}{\sqrt{2}}( m_{1z}  -m_{2z}   - m_{3z} + m_{4z})], \nonumber\\
M_{33} =& \f{3\sqrt{5}}{8} i[ m_{1x} - im_{1y} - m_{2x} + im_{2y} - m_{3x} + im_{3y} + m_{4x} - im_{4y}]. \label{eq:octu}
\end{align}
\end{widetext}

\subsection{Classification of CMMM by Irreducible Representations}

Multipole moments can be classified by irreducible representations (irreps) of symmetry group\cite{suzuki2017cluster,kusunose2008description,takimoto2006antiferro,kiss2005group,shiina1997magnetic,santini2009multipolar}. We can classify the CMMM in the same way.

Applying projection operators for CMMMs\cite{dresselhaus2007group}, we classify CMMM by irreps. Symmetrized CMMM can be considered as order parameters, since symmetrized CMMM represent the degree of symmetry breaking. In TABLE \ref{tab:CMMM} and Fig. \ref{fig:CMMM}, we show symmetrized CMMM as the linear combination of CMMM and as a configuration of the magnetic moments in the lattice.

In order to analyze the symmetry properties of $J=3/2$ states at quadratic band crossing, let us concentrate only on the Cluster 1. Since there are 12 degrees of freedom in Cluster 1, we have 12 independent symmetrized CMMMs. However, up to octupole, there should be 15 (3+5+7) order parameters. In fact, 3 octupolar symmetrized CMMMs $T_x^1, T_y^1, T_z^1$ corresponds to the symmetrized quadrupoles $Q_{yz},Q_{zx}.Q_{xy}$. Thus, the number of independent CMMMs are just as same as the number of degrees of freedom.

However, in the presence of inversion symmetry, quadrupole must vanish due to its oddness under inversion, then only dipoles and octupoles can exist in pyrochlore iridates. We clearly prove the statement by adding the configuration of Cluster 2 to that of Cluster 1. For Cluster 2, the orientation of magnetic moment at each site is opposite to that in Cluster 1 only in the quadrupole order.

\begin{table*}[ht]
\centering
\begin{tabular*}{\textwidth}{c@{\hspace{120pt}}c@{\hspace{120pt}}c}
\hline \hline Multipole  & Irrep  & CMMM \\ \hline 
Dipole & $T_1$ $(\Gamma_4)$  & $N_x \equiv \f{M_{1\bar1}-M_{11}}{\sqrt{2}}$ \\
          &                              & $N_y \equiv \f{M_{1\bar1}+M_{11}}{\sqrt{2}i}$ \\
          &                              & $N_z \equiv M_{10}$ \\ \hline
Quadrupole & $E$ $(\Gamma_3)$ & $Q_{3z^2-r^2} \equiv M_{20} $\\
		 &				& $Q_{x^2-y^2} \equiv \f{1}{\sqrt{2}}(M_{22}+M_{2\bar2})$ \\
		 & $T_1$ $(\Gamma_4)$ & $Q_{yz} \equiv -\f{i}{\sqrt{2}}(M_{21}+M_{2\bar1})$ \\
		 & 				 & $Q_{zx} \equiv \f{1}{\sqrt{2}}(-M_{21}+M_{2\bar1})$ \\
		 &				 & $Q_{xy} \equiv \f{i}{\sqrt{2}}(M_{22}-M_{2\bar2})$ \\\hline
Octupole    & $A_2$ $(\Gamma_2)$ & $T_{xyz} \equiv \f{i}{\sqrt{2}}(M_{32}-M_{3\bar2})$ \\
		& $T_1$ $(\Gamma_4)$ & $T^1_x \equiv \f{1}{4}[\sqrt{5}(-M_{33}+M_{3\bar3})-\sqrt{3}(-M_{31}+M_{3\bar1})]$ \\
		&				   & $T^1_y \equiv \f{i}{4}[\sqrt{5}(M_{33}+M_{3\bar3})+\sqrt{3}(M_{31}+M_{3\bar1})]$ \\
		&				   & $T^1_z \equiv M_{30}$ \\
		 & $T_2$ $(\Gamma_5)$ & $T^2_{x}\equiv \f{1}{4}[\sqrt{5}(M_{31}-M_{3\bar1}) + \sqrt{3}(M_{33}-M_{3\bar3})]$ \\
         &				  & $T^2_{y}\equiv \f{-i}{4}[\sqrt{5}(M_{31}+M_{3\bar1})-  \sqrt{3}(M_{33}+M_{3\bar3})]$ \\	
		 &				  & $T^2_{z}\equiv \f{1}{\sqrt{2}}(M_{32}+M_{3\bar2})$ \\ \hline \hline
\end{tabular*}
\caption{CMMM are classified into the irreps of $T_d$ group. The table is very similar to CMMMs in the reference\cite{suzuki2017cluster}. We show two kinds of group representation in the second column; $T_1$, $E$, $A_2$, $T_2$ are for $T_d$ single group, and $\Gamma_i$ are for double group.}
\label{tab:CMMM}
\end{table*}

%******************************* Tight-binding Model ******************************%
\section{The Lattice Model \label{sec:App4}}

%******************************* The detailed information of Tight-binding model ******************************%
\subsection{Phase diagrams}

The tight-binding model Hamiltonian is $H_{TB}=H_0+H_U+H_Z$\cite{witczak2013pyrochlore}. First, 
\begin{align}
H_0 = & \sum_{\langle ij \rangle} c^\dagger_i(t_1+it_2\mathbf{d_{ij}}\cdot\V \sigma)c_j \n&+ \sum_{\langle \langle ij \rangle \rangle }c^\dagger_i(t'_1+i[t'_2\mathbf{R_{ij}}+t'_3\mathbf{D_{ij}}]\cdot\V \sigma)c_j,
\end{align}
where it describes the nearest and next-nearest neighbor hopping. Note that the hopping vectors are defined as
\begin{align}
\mathbf{d}_{ij} =& 2\mathbf A_{ij}\times\mathbf B_{ij},~\mathbf A_{ij} = \f{1}{2}(\mathbf b_i + \mathbf b_j) - \mathbf c, \n\mathbf B_{ij} =& \mathbf b_j - \mathbf b_i,~ \mathbf{R}_{ij} = \mathbf B_{ik}\times \mathbf B_{kj},\n \mathbf{D}_{ij} =& \mathbf d_{ik}\times \mathbf d_{kj} 
\end{align}
where $\mathbf b_i$ is the position of $i$-th atom in the unit cell, $\mathbf c$ is the position of the center of the unit cell. The hopping parameters are defined as
\begin{align}
t_1 =& \frac{130}{243}t_{oxy}+\frac{17}{324}t_\sigma-\frac{79}{243}t_\pi \nonumber \\
t_2 =& \frac{28}{243}t_{oxy}+\frac{15}{243}t_\sigma-\frac{40}{243}t_\pi \nonumber \\
t_1'=& \frac{233}{2916}t_\sigma'-\frac{407}{2187}t_\pi' \nonumber \\
t_2'=& \frac{1}{1458}t_\sigma'+\frac{220}{2187}t_\pi' \nonumber \\ 
t_3'=& \frac{17}{324}t_\sigma'+\frac{460}{2187}t_\pi'. \nonumber
\end{align}
where $t_{\sigma,\pi}'=\alpha t_{\sigma,\pi}$.

Second, the Hubbard repulsion Hamiltonian $H_U$ is
\begin{align}
H_U=U \sum_{Ri} n_{Ri\uparrow}n_{Ri\downarrow}, \label{eq:UU}
\end{align}
where $n_{Ris}$ is the number operator of iridium electrons whose effective angular momentum is $1/2$. We apply Hartree-Fock approximation to this Hubbard repulsion term.
\begin{align}
H_{U}^{MF} =&~ -U (\sum_{Ri}2 \langle \V m_{R,i} \rangle \cdot \V m_{R,i} - \langle \V m_{R,i} \rangle^2),\n \V m_{R,i} =&~ \f{1}{2N} \sum_{\alpha,\beta=\uparrow,\downarrow} c_{Ri\alpha}^\dagger \sigma_{\alpha,\beta}c_{Ri\beta}.
\end{align}
where $N$ is the total number of unit cells in the lattice.

Finally, we have Zeeman coupling for Ir electrons, whose effective angular momentum is $1/2$.
\begin{align}
H_Z = - \frac{1}{2}\sum_{Ris} c_{Ris}^\dagger(\vec H\cdot \vec \sigma_{ss'})c_{Ris'}. \label{eq:Zee}
\end{align}

We can add an additional interaction into this Hamiltonian, which couples rare-earth $f$-electrons to iridium $d$-electrons\cite{tian2016field}. Since $f$-electrons also have spins, we should consider Zeeman effect for $f$-electrons. The Hamiltonian is $H' = H + H_{fd} + H_z'$, where
\begin{align}
H_{fd} =&~ J_{fd} \sum_{\langle iJ\rangle} \sum_{\mu,\nu=x,y,z} \Lambda_{iJ}^{\mu\nu} \sigma_i^\mu \tau_J^\nu, \n
H_z' =&~ -\sum_I \gamma (\vec H \cdot \vec a_I) \tau_I^z.
\end{align}
Here, $J_{fd}$ is the coupling constant, $\tau_I^\mu$ are the rare-earth f-electron spins which are Ising-like along local [111] direction, $\gamma$ is f-electron g-factor, and $\Lambda_{iJ}^{\mu\nu}$ are defined\cite{tian2016field} as
\begin{align}
\Lambda_{iJ}^{\mu\nu} = \left\langle\begin{matrix}
[G_1^x \V a_J + G_2^x \V a_J \bar{\times}(\V d_{iJ} \bar\times \V d_{iJ})]\cdot \hat e_\mu, ~ (\nu=x)\\ G^y \V a_J \times (\V d_{iJ} \bar{\times} \V d_{iJ}) \cdot \hat e_\mu,~ (\nu=y)\\ [G_1^z \V a_J + G_2^z \V a_J \bar{\times}(\V d_{iJ} \bar\times \V d_{iJ})]\cdot \hat e_\mu,~ (\nu=z)
\end{matrix}\right]. \nonumber
\end{align}
for Nd$^{3+}$, which is a Kramers ion. Here, $i,j$ are for iridium site while $I,J$ are for rare-earth site. However, for Pr$^{3+}$, which is a non-Kramers ion,
\begin{align}
\Lambda_{iJ}^{\mu\nu} = \left\langle\begin{matrix}
0,  ~ (\nu=x)\\ 0, ~ (\nu=y)\\ [G_1^z \V a_J + G_2^z \V a_J \bar{\times}(\V d_{iJ} \bar\times \V d_{iJ})]\cdot \hat e_\mu,~ (\nu=z) 
\end{matrix}\right].\nonumber
\end{align}
Furthermore, Pr in-plane components can couple to the charge density of Ir electrons\cite{lee2013RKKY}.

\begin{figure}
\centering
\includegraphics[width=0.75\columnwidth]{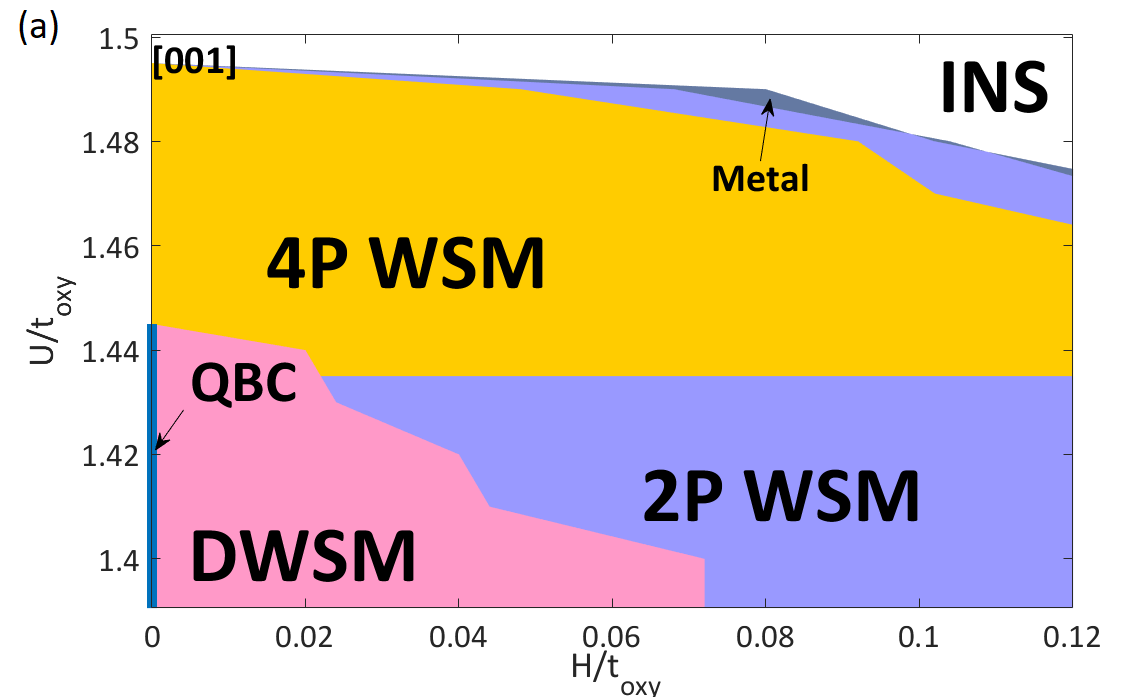}
\includegraphics[width=0.75\columnwidth]{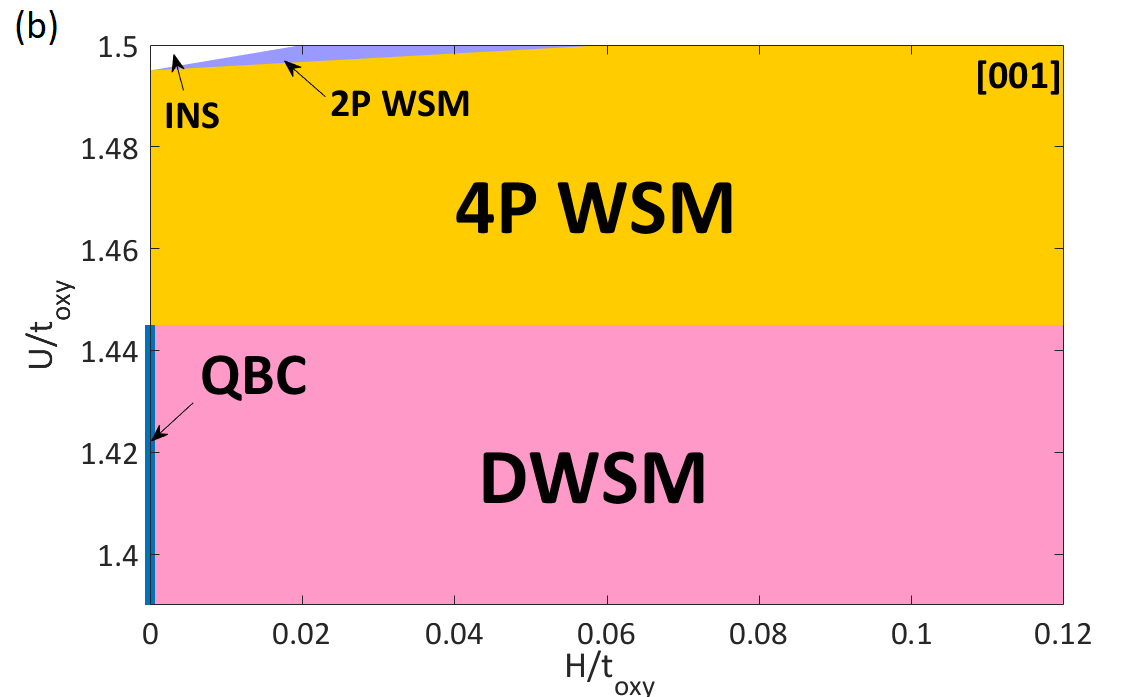}
\includegraphics[width=0.75\columnwidth]{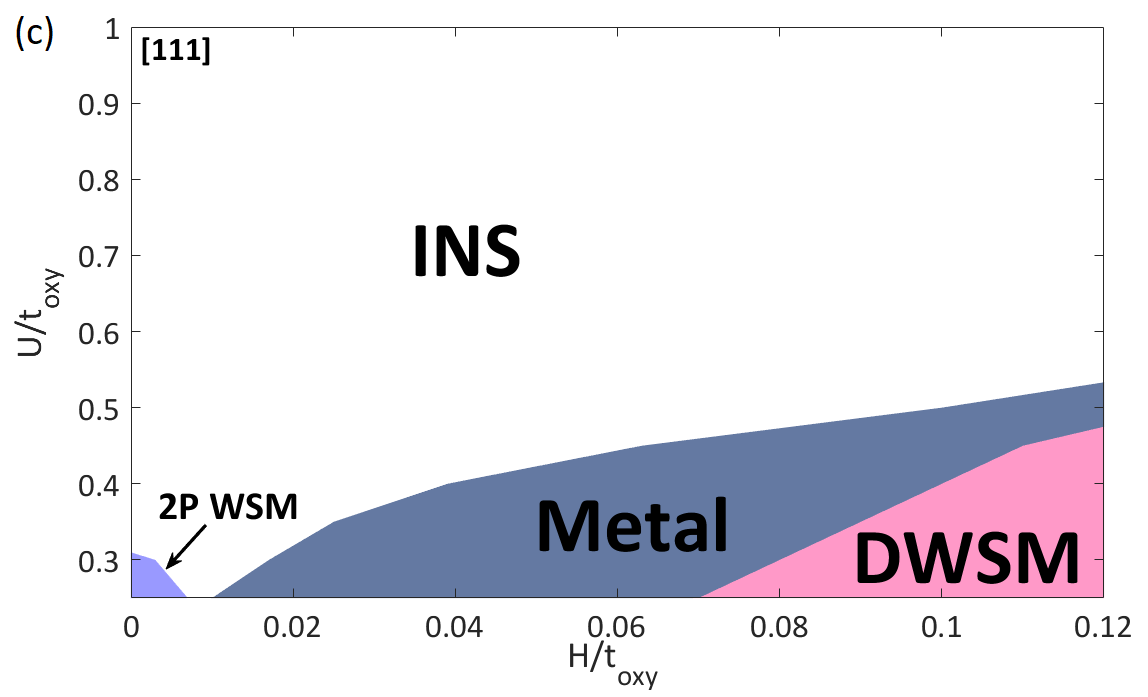}
\caption{General phase diagrams from self-consistent mean-field theory. (a) $t_{\sigma}=-0.8, \alpha=0.08, J_{fd}=0,$ $[001]$ field. Heisenberg Ir spin. (b) $t_{\sigma}=-0.8, \alpha=0.08, J_{fd}=0,$ $[001]$ field. Ising Ir spin. (c) $t_{\sigma}=-1.1, \alpha=0.02, J_{fd}=1,$ $[111]$ field. Heisenberg Ir spin.}
\label{fig5.phase}
\end{figure}

We obtain the ground state energy band and magnetic moment configuration through self-consistent mean-field theory under various Hubbard strength $U$ and magnetic field strength $H$ for Nd$_2$Ir$_2$O$_7$. Then, we investigate crossing points within Brillouin zone to determine topological phases, and exhibit the general phase diagrams with different parameters in Fig.~\ref{fig:7} and~\ref{fig5.phase}.

%******************************* Projection onto effective theory ******************************%

\subsection{Projection of the Effective Zeeman Field onto the Effective Theory}

We can consider Hubbard repulsion, $fd$-exchange, and the magnetic field altogether in the effective field which is applied for each iridium spin. Then the interaction Hamiltonian is just an effective Zeeman term,
\begin{align}
H_B =& H_U^{MF}+H_{fd}+H_Z 
\n=& \f{1}{2}\sum_{i} \vec{B}_{\text{eff},i} \cdot [c^\dagger_{i,s} \vec{\sigma}_{ss'} c_{i,s'}] + \text{const}. ,
\end{align}
where 
\begin{align}
\f{1}{2} B_{\text{eff},i}^\mu =  - \f{U}{2}\langle \sigma_i^\mu \rangle + J_{fd} \sum_{J,\nu} \Lambda_{iJ}^{\mu\nu}\tau_J^{\nu} - \f{1}{2}H^{\mu}.
\end{align}

Since the magnetic moment has the same symmetric properties as effective field, we can define symmetrized CMMMs in terms of effective field instead of magnetic moments. That is, the magnetic moments in Eq.~\ref{eq:dipo},~\ref{eq:quad},~\ref{eq:octu} are just replaced with $\vec{B}_{\text{eff},i}$. After then, let us define some order parameters with effective field based symmetrized CMMMs. AIAO order parameter is defined as
\begin{align}
M_{A_2} = \frac{1}{8}\sum_i \vec{B}_{\text{eff},i} \cdot \vec{a_i} = \f{1}{2\sqrt{15}} T_{xyz},
\end{align}
such that $\vec{a_i}$ is the unit vector directing from the $i$-th site to the center of tetrahedron. This changes as $\Gamma_2$ representation of $T_d$ double group. The magnetization is defined as
\begin{align}
M_{D,\mu} = \f{1}{8}\sum_{i} B_{\text{eff},i\mu} = \f{1}{8}N_\mu,
\end{align}
where $\mu = x,y,z$. 2I2O order parameter is defined by $T_1$ octupole $(T_x^1,T_y^1,T_z^1)$,
\begin{widetext}
\begin{align}
M_{T_1,x} =& \f{1}{12}(B_{\text{eff},1y}+B_{\text{eff},1z}-B_{\text{eff},2y}-B_{\text{eff},2z}-B_{\text{eff},3y}+B_{\text{eff},3z}+B_{\text{eff},4y}-B_{\text{eff},4z}) = -\f{1}{9}T_x^1\n
M_{T_1,y} =& \f{1}{12}(B_{\text{eff},1x}+B_{\text{eff},1z}-B_{\text{eff},2x}+B_{\text{eff},2z}-B_{\text{eff},3x}-B_{\text{eff},3z}+B_{\text{eff},4x}-B_{\text{eff},4z}) = -\f{1}{9}T_y^1\n 
M_{T_1,z} =& \f{1}{12}(B_{\text{eff},1x}+B_{\text{eff},1y}-B_{\text{eff},2x}+B_{\text{eff},2y}+B_{\text{eff},3x}-B_{\text{eff},3y}-B_{\text{eff},4x}-B_{\text{eff},4y}) = -\f{1}{9}T_z^1.
\end{align}
\end{widetext}
Those order parameters commonly appear for both $[001]$ and $[111]$ direction field. 

For the projection of the lattice model, we find $J_z$ eigenstates from taking fourfold degenerate eigenstates of $H_0$ at $\Gamma$ point (Fig. 6(a)). $J_z$ eigenstates  $\{\ket{\psi_{j_z}}\}$  are
\begin{align}
\ket{\psi_{3/2}} =&~ \f{1}{2}(e^{i\f{3\pi}{4}},e^{i\f{\pi}{4}},e^{-i\f{3\pi}{4}},e^{-i\f{\pi}{4}},0,0,0,0)^T \n
\ket{\psi_{1/2}} =&~ \f{1}{\sqrt{6}}(-i,i,i,-i,\f{-1+i}{2},\f{1+i}{2},\f{-1-i}{2},\f{1-i}{2})^T \nonumber
\end{align}
\begin{align}
\ket{\psi_{-1/2}}=&~ \f{1}{\sqrt{6}}(\f{-1-i}{2},\f{1-i}{2},\f{-1+i}{2},\f{1+i}{2},-i,i,i,-i)^T \n
\ket{\psi_{-3/2}}=&~ \f{1}{2}(0,0,0,0,e^{-i\f{3\pi}{4}},e^{-i\f{\pi}{4}},e^{i\f{3\pi}{4}},e^{i\f{\pi}{4}})^T 
\end{align}
Then, the projection matrix is just $P = \sum_{j_z} \ket{\psi_{j_z}}\bra{\psi_{j_z}}$.

We have total 12 degrees of freedom (4 site $\times$ 3 directions), but we can reduce the number of parameter into 4 by symmetry. Considering $C_{2z}$ and $\sigma_d T$, the symmetries under $[001]$ magnetic field and AIAO order, we have in general,
\begin{align}
\vec{B}_{\text{eff},1} =& (B_{\text{eff},1x},B_{\text{eff},1x},B_{\text{eff},1z}) \n
\vec{B}_{\text{eff},2} =& (B_{\text{eff},2x},-B_{\text{eff},2x},B_{\text{eff},2z}) \n
\vec{B}_{\text{eff},3} =& (-B_{\text{eff},2x},B_{\text{eff},2x},B_{\text{eff},2z}) \n
\vec{B}_{\text{eff},4} =& (-B_{\text{eff},1x},-B_{\text{eff},1x},B_{\text{eff},1z}) 
\end{align}
Under the magnetic moment configuration, the order parameters are
\begin{align}
M_{A_2} =& \f{1}{8\sqrt{3}}(4B_{\text{eff},1x} + 2B_{\text{eff},1z} + 4B_{\text{eff},2x} - 2B_{\text{eff},2z}) \n
M_{D,z} =& \f{1}{4}(B_{\text{eff},1z}+B_{\text{eff},2z}) \n
M_{T_1,z}=& \f{1}{3}(B_{\text{eff},1x}-B_{\text{eff},2x}). 
\end{align}

We can now express the projection of the effective Zeeman term as
\begin{align}
P^\dagger H_B P =& (M_{A_2} \Gamma_{45} + (\f{2}{3} M_{D,z} -\f{9}{4} M_{T_1,z} ) J_z - M_{T_1,z} J_z^3).
\end{align}

On the other hand, if we consider $C_{3,[111]}$ and $\sigma_{d,[111]} T$, the symmetries under $[111]$ magnetic field and AIAO order, we can reduce the number of parameter into 3 by symmetry.
\begin{align}
\vec{B}_{\text{eff},1} =& (B_{\text{eff},1x},B_{\text{eff},1x},B_{\text{eff},1x}) \n
\vec{B}_{\text{eff},2} =& (B_{\text{eff},2x},B_{\text{eff},2y},B_{\text{eff},2y}) \n
\vec{B}_{\text{eff},3} =& (B_{\text{eff},2y},B_{\text{eff},2x},B_{\text{eff},2y}) \n
\vec{B}_{\text{eff},4} =& (B_{\text{eff},2y},B_{\text{eff},2y},B_{\text{eff},2x}) 
\end{align}
In this configuration, the order parameters are
\begin{align}
M_{A_2} =& \f{\sqrt{3}}{8}(B_{\text{eff},1x}+B_{\text{eff},2x}-2B_{\text{eff},2y}) \n
M_{D,x} =& M_{D,y} = M_{D,z} = \f{1}{8}(B_{\text{eff},1x}+B_{\text{eff},2x}+2B_{\text{eff},2y})\n
M_{T_1,x} =& M_{T_1,y} = M_{T_1,z} = \f{1}{6}(B_{\text{eff},1x}-B_{\text{eff},2x})
\end{align}
The projection of the effective Zeeman term under $[111]$ field is
\begin{align}
P^\dagger H_B P =& (M_{A_2} \Gamma_{45} + (\f{2}{3} \vec M_{D} -\f{9}{4} \vec M_{T_1} ) \cdot \vec J \n&- \vec M_{T_1} \cdot \vec J^3).
\end{align}

For both cases, we obtain $\theta$ and $\phi$,
\begin{align}
\theta =& \arctan\f{M_{T_1,z}}{\f{2}{3}M_{D,z} - \f{9}{4}M_{T_1,z}} \n
\phi =& \arctan\f{\f{2}{3}M_{D,z}-\f{9}{4}M_{T_1,z}}{M_{A_2} \cos\theta}.
\end{align}

%* *****************************************************  Bibliography ****************************************************** %

\end{document}